\documentclass[showpacs,preprintnumbers,amsmath,amssymb, pr, twocolumn]{revtex4-1}

\usepackage{stmaryrd}
\usepackage{txfonts}
\usepackage{amssymb}
\usepackage{mathrsfs}
\usepackage{graphicx}
\usepackage{dcolumn}
\usepackage{bm}
\usepackage{epsfig}
\begin{document}

\preprint{S. -Z. Lin, Phys. Rev. B {\bf{86}}, 014510 (2012)}

\title{Josephson effect between a two-band superconductor with the $s++$ or $s\pm$ pairing symmetry and a conventional $s$-wave superconductor}

\author{Shi-Zeng Lin}
\email{szl@lanl.gov}

\affiliation{Theoretical Division, Los Alamos National Laboratory, Los Alamos, New Mexico 87545, USA}

\begin{abstract}
In this work, we investigate the Josephson effect between a two-band superconductor either with the $s++$ (two energy gaps have the same sign and are fully gapped) pairing symmetry or $s\pm$ (two energy gaps have $\pi$ phase difference and are fully gapped) pairing symmetry and a conventional $s$-wave superconductors. The ground state, critical current, plasma modes, flux flow dynamics, and response to external ac electric field, possible soliton solutions are investigated. For junctions with the charge neutrality breaking, we find a new plasma mode for junctions, which gives rise to new resonance peaks in the Josephson flux flow region. Because of the frustrated interaction in junctions with $s\pm$ pairing symmetry, time-reversal-symmetry (TRS) can be broken if the frustration is optimized. In the TRS broken (TRSB) state, there is a non-trivial phase difference between the two Josephson tunnelling channels, which results in a non-trivial interference. Furthermore, we find a novel massless plasma mode at the TRSB transition for junctions with the charge neutrality breaking. In the TRSB state, a spontaneous magnetic flux appears where there is a spatial inhomogeneity in the Josephson coupling, thus provides a possible smoking-gun evidence for the underlying pairing symmetry.     
\end{abstract}


\date{\today}

\maketitle
\section{Introduction}
The Josephson effect between two superconductors is a hallmark of the macroscopic quantum effect associated with superconductivity\cite{Josephson62}. When a phase difference exists between two conventional superconductors, spontaneous supercurrent flows from one superconductor to the other when they are brought together to form a Josephson junction. This is the celebrated dc Josephson effect which describes the relation between the current and phase difference $I_s=I_c \sin(\theta_1-\theta_2)$, where $I_c$ is the critical current and $\theta_i$ is the superconducting phase. When a voltage $V$ is applied to the junction, the superconducting phase rotates according to the ac Josephson effect $\hbar\partial_t(\theta_1-\theta_2)=2eV$. Because of the novel quantum nature of the Josephson effect, Josephson junctions have wide applications, such as SQUID, electromagnetic devices and quantum qubit\cite{BaroneBook}.

The Josephson effect involves the phase of the superconductivity therefore depends sensitively on the underlying pairing symmetry of the superconductors. Thus the Josephson effect provides an invaluable tool to pin down the paring symmetry of some exotic superconductors, such as Geshkenbein, Larkin and Barone phase sensitive measurement for $p$-wave pairing symmetry in heavy-fermion superconductors\cite{Geshkenbein87}, and tri-junction in cuprate superconductors\cite{Tsuei00}. On the other hand, the current-phase relation is modified in these exotic junctions\cite{Golubov04}, which broadens our general understanding on the Josephson effect and also points a new possible direction for applications.
  
The multiband superconductivity attracts considerable interests since the discoveries of $\text{MgB}_2$ superconductor\cite{Nagamatsu01} and iron pnictide superconductors\cite{Kamihara08}. The pairing symmetry in $\text{MgB}_2$ now is well-understood. It has two $s$-wave energy gaps with the same phase in the $\sigma$ and $\pi$ bands, known as $s++$ pairing symmetry\cite{Xi08}. While for pnictide superconductors, due to vastness of iron pnictide family, discrepancies from different measurements and controversies remain, see Ref.\cite{Ishida09,Paglione10,Johnston10,Wen11,Wang11,Hirschfeld11,Stewart11} for a review. The so-called extended $s$-wave or $s\pm$ pairing symmetry with a $\pi$ phase shift between the hole and electron Fermi surface\cite{Mazin08,Kuroki08} has attracted considerable attention and is favoured by many experiments\cite{Ding08,Wray08,Terashima09,Zhang10b}, and is considered as a most probable candidate for iron pnictide superconductors.

\begin{figure*}[t]
\psfig{figure=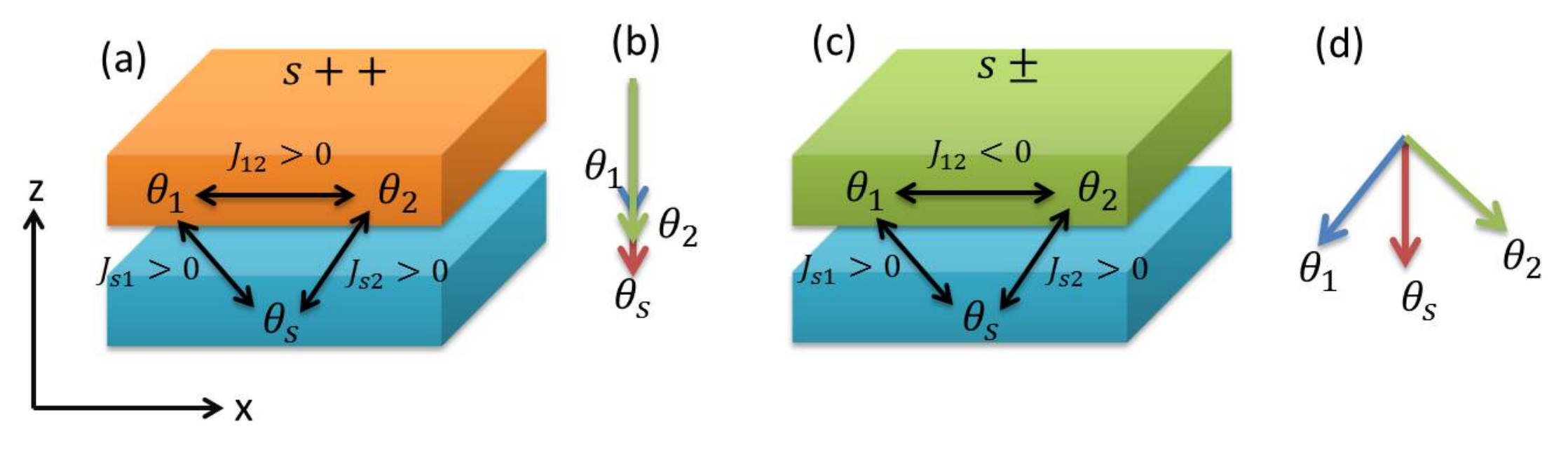,width=16cm} \caption{\label{f1} (color online). (a) Josephson junctions between two-band superconductors with the $s++$ pairing symmetry and $s$-wave single band superconductors. The interaction between condensates is attractive and they have the same phase in the ground state as shown in (b). (c) Josephson junctions between two-band superconductors with the $s\pm$ pairing symmetry and $s$-wave single band superconductors. The interaction between condensates $\theta_1$ and $\theta_2$ is repulsive. Under appropriate conditions, the system is strongly frustrated resulting in TRSB and the ground state is depicted in (d).}
\end{figure*}

Soon after the discoveries of iron pnictide superconductors, Josephson junctions have been fabricated between a conventional superconductor and these multiband superconductors, with an aim to determine the pairing symmetry of pnictide. Zhang \emph{et. al.}\cite{Zhang09} fabricated junction between BaKFeAs and a conventional superconductor Pb, they clearly observed the Fraunhofer pattern in the dependence of the critical current on magnetic fields and Shapiro steps when the junction is irradiated by microwaves. These observations indicate that a pure $p$-wave or $d$-wave pairing symmetry is unlikely realized in pnictide. Similar Fraunhofer pattern is observed by Zhou \emph{et. al.} in a corner junction between BaFeCoAs and Pb\cite{Zhou08}. Recently, unambiguous half-integer flux quantum jumps have been observed in polycrystal NdFesO-Niobium superconducting loop\cite{Chen10}, which is a strong evidence for the $s\pm$ pairing symmetry. For a review on $\text{MgB}_2$ junctions and all pnictide junctions, please see Ref. \cite{Brinkman03,Xi09,Seidel11}

From a theoretical point view, for Josephson junctions between two-band superconductors and conventional $s$-wave superconductors, there are two superconducting tunnelling channels $J_{s1}\sin(\theta_1-\theta_s)$ and    $J_{s2}\sin(\theta_2-\theta_s)$, where $\theta_s$ is the phase of the $s$-wave superconductor and $\theta_i$ with $i=1,2$ is the phase of the two-band superconductor. The relative phase between $\theta_1$ and $\theta_2$ is fixed by the underlying pairing symmetry. For the $s++$ pairing symmetry, such as $\text{MgB}_2$, these two channels have the same phase and add constructively, thus the phase-current relation is equivalent to a conventional junction made of single band superconductors with an effective Josephson coupling $J_{\text{eff}}=J_{s1}+J_{s2}$. While for the $s\pm$ pairing symmetry, non-trivial phenomena unique to the sign-reversal pairing symmetry arise. It was discussed by Agterberg \emph{et. al.}\cite{Agterberg02} even before the discovery of pnictide superconductors that the critical current depends non-monotonically on temperatures, and may even becomes negative in some temperature region. Assuming a $\pi$ phase shift between the two tunnelling channels, numerous phenomena have been demonstrated such as, realization of the $\pi$ junction\cite{Chen09,Parker09}, vortex enlargement\cite{Ota10}, new Shapiro steps\cite{Ota10b}, upper bound of the critical current from the Ambegaokar - Baratoff relation\cite{Ota10c}. We note that the most studies only show quantitatively difference between junctions with $s++$ and $s\pm$ pairing symmetry.

Due to the $s\pm$ pairing symmetry, the system is somewhat frustrated as shown in Fig. \ref{f1}(d). Non-trivial phase difference between two tunnelling channels beside $0$ with an effective Josephson coupling $J_{s1}+J_{s2}$ and $\pi$ with an effective Josephson coupling $J_{s1}-J_{s2}$ is possible. If this happens, time-reversal symmetry (TRS) is broken and there are two degenerate ground states. The possible violation of TRS is first discussed in Ref. \cite{Ng09} and later in Ref.\cite{Stanev10} in the context of three-band superconductors. As a consequence of time-reversal symmetry breaking (TRSB), qualitatively different behaviour between junctions made of $s\pm$ and $s++$ superconductors may exist, which is hopeful to resolve the dispute of the pairing symmetry in pnictide superconductors by observing the differences in experiments.

In this work, we investigate the Josephson effect between a two-band superconductor with either $s++$ or $s\pm$ pairing symmetry and a conventional s-wave superconductor, with an emphasis on the possible TRSB. First we derive equations of motion for the gauge invariant phase differences based on the Lagrangian approach. Then the ground state is obtained by minimizing the Josephson energy. We find for junctions with $s\pm$ pairing symmetry under appropriate conditions, TRSB occurs. As a consequence of the TRSB, the critical current and height of the Shapiro step when junction is shined by an ac electric field develop a non-trivial phase dependence on the Josephson coupling. We proceed to investigate the Josephson plasma mode in junctions. For thin superconducting electrodes, charge neutrality breaking occurs resulting in a modified ac Josephson relation. A new plasma dispersion associated with an out-of-phase oscillation of phase difference emerges. At the TRSB transition, this new plasma mode becomes massless. Because of the existence of the massless mode, the Josephson penetration depth diverges and the lower critical field $H_{c1}$ of the junction vanishes. In the flux flow region, we obtain additional Fiske resonances for a given cavity index and an additional Eck resonance because of the charge neutrality breaking. 

In the last part of the paper, we discuss possible topological excitations in the junction because of the existence of multiple degenerate energy minima. In the TRSB state, a new type of soliton can be stabilized between the TRSB pair states. Finally we discuss appearance of spontaneous magnetic flux when there is a spatial variation of the Josephson coupling, which occurs only in the junction with $s\pm$ pairing symmetry, thus provides a possible smoking-gun evidence for the pairing symmetry.        

The remaining part of the paper is organized as follows: in Sec. II, we derive equations of motion for the gauge-invariant phase differences. In Sec. III, the boundary condition is derived. In Sec. IV, we calculate the phase differences at the ground state. In Sec. V, the dependence of the critical current on magnetic fields is obtained. In Sec. VI, the response of the junction to an incident electromagnetic wave is investigated. In Sec. VII, we calculate the Josephson plasma modes in the junction. In Sec. VIII, the Josephson penetration depth and the lower critical field are studied. In Sec. IX, the McCumber solution is obtained. In Sec. X, we investigate the Fiske and Eck resonances in the flux flow region. In Sec. XI, we discuss the topological excitations (soliton) in the junction. In Sec. XII, possible spontaneous magnetic flux is investigated for junctions made of superconductors with $s\pm$ pairing symmetry. The paper is closed by conclusions in Sec. XIII.

\section{Model}
We consider a junction between a conventional s-wave superconductor and a multi-band superconductor either with $s++$ or $s\pm$ pairing symmetry, as depicted in Fig. \ref{f1}. The analysis can be extended straightforwardly to junctions between two two-band superconductors. However the physics is expected to be qualitatively the same as the heterotic junctions studied here. In the junction between a $s++$ superconductor and a $s$-wave superconductor, the Josephson couplings among different condensates are positive, thus in the ground state they have the same phase. While for the junction between a $s\pm$ superconductor and a $s$-wave superconductor, the inter-band Josephson coupling between condensates with phase $\theta_1$ and $\theta_2$ in the $s\pm$ superconductor is negative, and the inter-junction Josephson coupling between the $s$-wave superconductor and $s\pm$ superconductor is positive, which results in frustration in the system. When the frustration is optimized, different condensates have non-zero phase differences among them which indicates the breaking of TRS. We will show below that the behavior is qualitatively different for junctions in Fig. \ref{f1}(a) with TRS and Fig. \ref{f1}(c) with TRSB, which implies a possible way to detect the pairing symmetry for multiband superconductors. 

The Hamiltonian of the junction is
\begin{equation}\label{eqH1}
\mathcal{H}=\mathcal{H}_{1}+\mathcal{H}_{2}+\mathcal{H}_{t},
\end{equation}
with the Hamiltonian for the single band superconductor
\begin{equation}\label{eqH2}
\mathcal{H}_1=\int {d^3}r\sum\limits_{\sigma } { c _{\sigma }^\dag (\mathbf{r}){(\varepsilon_{c}-\mu_c)} {c _{\sigma }}(\mathbf{r}) }- g c _{\sigma }^\dag (\mathbf{r}) c_{\bar{\sigma} }^\dag (\mathbf{r}){c _{\bar{\sigma} }}(\mathbf{r}){c _{\sigma }}(\mathbf{r})
\end{equation}
and the Hamiltonian for the two-band superconductor
\begin{align}\label{eqH3}
\nonumber \mathcal{H}_2= \sum\limits_{l,\sigma } {\int {{d^3}rd _{l\sigma }^\dag (\mathbf{r}){(\varepsilon_{d,l}-\mu_d)} {d _{l\sigma }}(\mathbf{r})} } \\
  - \sum\limits_{j,l=1,2} {\int } {d^3}r d _{j\sigma }^\dag (\mathbf{r}) d_{j\bar{\sigma} }^\dag (\mathbf{r}){V_{jl}}{d _{l \bar{\sigma} }}(\mathbf{r}){d _{l\sigma }}(\mathbf{r}),
\end{align}
and the tunnelling between the two superconductors
\begin{equation}\label{eqH4}
\mathcal{H}_t=\sum_{l,\sigma}t_{l,s}c _{\sigma }^\dag  d _{l\sigma }+\rm{H. C.},
\end{equation}
where $d _{l\sigma }^\dag$ (${d _{l\sigma }}$) is the electron
creation (annihilation) operator in the $l$-th band of the two-band superconductor with the
dispersion $\varepsilon_{d, l}(\mathbf{k})$ and the chemical potential
$\mu_d$ and spin index $\sigma$. $V_{jl}$ is the intra-band for $l=j$
and inter-band for $l\neq j$ scattering respectively, which can be
either repulsive or attractive depending, for instance, on
the strength of the Coulomb and electron-phonon interaction. $c _{\sigma }^\dag$ is the electron creation operator in the single band superconductor and $g$ is the electron-phonon coupling strength. $t_{l,s}$ is the tunnelling matrix for electrons between the two superconductors.

A schematic view of the junction geometry is depicted in Fig. \ref{f1}. External magnetic fields are applied along the $y$ direction, which induces inhomogeneous phase differences along the $x$ direction. We assume the system is uniform along the $y$ direction and the problem becomes two dimensional. We proceed to derive equations of motion for the gauge invariant phase differences for the junction using the Lagrangian approach \cite{Simanek94,Ota09}. The total Lagrangian of system has three contributions
\begin{equation}\label{eq1}
\mathcal{L}=\mathcal{L}_{1}+\mathcal{L}_{2}+\mathcal{L}_{B}
\end{equation}
with the Lagrangian for the single band superconductor, which can be derived from the Hamiltonian Eq. (\ref{eqH2}) using a standard method\cite{SimonsQFT}
\begin{align}\label{eq2}
\nonumber\mathcal{L}_1=\frac{d}{8\pi \mu _s^2}\left[A_0^B(r)+\frac{\Phi _0}{2\pi  c}\partial _t\theta _s(r,t)\right]^2\\
-\frac{d}{8\pi \lambda _s^2}\left[A_x^B(r)-\frac{\Phi _0}{2\pi  }\nabla \theta _s(r,t)\right]^2
\end{align}
and the Lagrangian the two-band superconductor
\begin{align}\label{eq3}
\nonumber{{\cal L}_{2}} = \sum\limits_{i = 1,2} {\frac{d}{{8\pi {\mu _i^2}}}{{\left[ {A_0^T(r) + \frac{{{\Phi _0}}}{{2\pi c}}{\partial _t}{\theta _i}(r,t)} \right]}^2}} \\
 - \sum\limits_{i = 1,2} {\frac{d}{{8\pi {\lambda _i^2}}}{{\left[ {A_x^T(r) - \frac{{{\Phi _0}}}{{2\pi }}{\partial _x}{\theta _i}(r,t)} \right]}^2}} +\frac{J_{12}\Phi_0}{2\pi c} \cos(\theta_1-\theta_2)
\end{align}
where $\mu_s$ and $\mu_i$ are the Thomas-Fermi lengths associated with charge screening, and $\lambda_s$ and $\lambda_i$ are the penetration depths for each band, and $\theta_s$ and $\theta_i$ are the superconducting phases for different condensates respectively. The effective penetration depth for the two-band superconductor is $\lambda_L^{-2}=\lambda_1^{-2}+\lambda_2^{-2}$. $J_{12}$ is the inter-band Josephson coupling, and $J_{12}<0$ for the $s\pm$ pairing symmetry while $J_{12}>0$ for the $s++$ pairing symmetry. $A_0^B$ and $A_0^T$ are electric potentials, and $A_x^B$ and $A_x^T$ are vector potentials at the bottom and top electrodes. $\Phi_0=h c/2e$ is the quantum flux. Here we have introduced the charge energy [the first term at the right-hand side of Eqs. (\ref{eq2}) and (\ref{eq3})] in the superconductors to account for the possible charge neutrality breaking\cite{Suhl65,Koyama96,Machida99} when the thickness of superconducting electrodes $d$ is comparable to $\mu_s$ or $\mu_i$, which might be realized in layered superconductors with strong anisotropy\cite{Kleiner92}.

The Lagrangian for the insulating barrier reads
\begin{equation}\label{eq4}
\mathcal{L}_{B}=\frac{b \epsilon _b}{8\pi }E_{b,z}^{2}-\frac{b}{8\pi }B_{b,y}^{2}-V_J,
\end{equation}
with $b$ being the thickness of the barrier and $\epsilon_d$ the dielectric constant. The electric field in the barrier is
\begin{equation}\label{eq5}
E_{b,z}=-\frac{1}{c}\partial _tA_{b,z}-\partial _zA_0=-\frac{1}{c}\partial _tA_{b,z}-\frac{A_0^T-A_0^B}{b},
\end{equation}
and the magnetic field is
\begin{equation}\label{eq6}
B_{b,y}=\partial _zA_x-\partial _xA_{b,z}=\frac{A_x^T-A_x^B}{b}-\partial _xA_{b,z}.
\end{equation}
The Josephson coupling $V_J$ is
\begin{equation}\label{eq7}
V_J=-\frac{J_{s1}\Phi_0}{2\pi c}\cos(\phi_{s1})-\frac{J_{s2}\Phi_0}{2\pi c}\cos(\phi_{s2}),
\end{equation}
with the gauge invariant phase difference
\begin{equation}\label{eq8}
\phi_{si}\equiv \theta _i-\theta _s-\frac{2\pi  b}{\Phi _0}A_{b, z},
\end{equation}
with $i=1,\ 2$ and $\theta_1-\theta_2=\phi_{s1}-\phi_{s2}$. Here $J_{si}>0$ are the Josephson couplings and can be derived from a microscopic theory\cite{BaroneBook}
\begin{equation}\label{eq9}
J_{si}=\frac{2\hbar}{e R_{bi}}\frac{|\Delta_i\Delta_s|}{|\Delta_s|+|\Delta_i|}K\left(\frac{|\Delta_i|-|\Delta_s|}{|\Delta_i|+|\Delta_s|}\right),
\end{equation}
for a temperature much smaller than the critical temperature, where $K(x)$ is the complete eliptic integral of the first kind, and $\Delta_{s}$, $\Delta_{i}$ are the superconducting energy gaps for different condensates. The resistance for each channel at the barrier is
\begin{equation}\label{eq9a}
R_{bi}=\frac{\hbar^3}{4\pi}\frac{1}{e^2 N_{i}(0)N_{s}(0)t_{i,s}},
\end{equation}
where $N_i(0)$ is the density of state of quasiparticles in each band for the two-band superconductor and $N_s(0)$ is the density of state for the single band superconductor. The Josephson couplings depend on the resistance of the barrier for each channel $R_{bi}$ and temperature, which implies a practical way of tuning in experiments\citep{,Agterberg02,Linder09}. 
 
For convenience, we introduce a length $\lambda_{c1}=\sqrt{c\Phi_0/(8\pi^2b J_{s1})}$ and a frequency $\omega_{p1}=c/(\sqrt{\epsilon_d} \lambda_{c1})$. We then renormalize the length in unit of $\lambda_{c1}$ and time in unit of $1/\omega_{p1}$. Current density is in unit of $J_{s1}$, electric field is in unit of $\Phi_0\omega_{p1}/(2\pi c b)$, and magnetic field is in unit of $\Phi_0/(2\pi \lambda_{c1} b)$.

We then minimize the Lagrangian in Eq. (\ref{eq1}) using the Euler-Lagrange equation. Applying the Euler-Lagrange equation with respect to $A_0^T$ and $A_0^B$, we have
\begin{equation}\label{eq10}
\sum _{i=1,2}\frac{1}{\alpha _i}\left[A_0^T+\partial _t\theta _i\right]-\epsilon _b\left[-b\partial _tA_b^z-\left(A_0^T-A_0^B\right)\right]=0,
\end{equation}
\begin{equation}\label{eq11}
\frac{1}{\alpha _s}\left[A_0^B+\partial _t\theta _s\right]+\epsilon _b\left[-b\partial _tA_b^z-\left(A_0^T-A_0^B\right)\right]=0,
\end{equation}
where $\alpha_{s(i)}\equiv u_{s(i)}^2/(db)$ are parameters characterizing the importance of the charge neutrality breaking. Equations (\ref{eq10}) and (\ref{eq11}) can be combined into the following equation using Eqs. (\ref{eq5}) and (\ref{eq8})
\begin{equation}\label{eq12}
\sum _{i=1,2}\frac{\partial _t\phi _{\text{si}}}{\alpha _i}=\left[\left(1+\frac{\alpha _s}{\alpha _1}+\frac{\alpha _s}{\alpha _2}\right)\epsilon _b+\frac{1}{\alpha _1}+\frac{1}{\alpha _2}\right]E_{b,z}\equiv C_eE_{b,z}
\end{equation}
Equation (\ref{eq12}) is a modified ac Josephson relation as a result of charge neutrality breaking. In the limit $\alpha_i\rightarrow 0$, we recover the standard ac Josephson relation $\partial_t \phi_{si}=E_{b,z}$. We will show that a new plasma mode emerges for non-zero $\alpha_i$ and $\alpha_s$.

Similarly minimizing the Lagrangian with respect to $A_x^T$ and $A_x^B$, we have
\begin{equation}\label{eq13}
-\sum _{i=1,2}\frac{1}{\zeta _i}\left[A_x^T-\partial _x\theta _i\right]-\left(A_x^T-A_x^B-b\partial _xA_z\right)=0,
\end{equation}
\begin{equation}\label{eq14}
-\frac{1}{\zeta _s}\left[A_x^B-\partial _x\theta _s\right]+\left(A_x^T-A_x^B-b\partial _xA_z\right)=0,
\end{equation}
where $\zeta_{s(i)}\equiv \lambda_{s(i)}^2/(db)$. We then have the relation between the magnetic field and spatial derivative of the gauge invariant phase difference after rewriting Eqs. (\ref{eq13}) and (\ref{eq14}) using Eqs. (\ref{eq6}) and (\ref{eq8})
\begin{equation}\label{eq15}
\sum _{i=1,2}\frac{\partial _x\phi _{\text{si}}}{\zeta _i}=\left[\left(1+\frac{\zeta _s}{\zeta _1}+\frac{\zeta _s}{\zeta _2}\right)+\frac{1}{\zeta _1}+\frac{1}{\zeta _2}\right]B_{b,y}\equiv C_bB_{b,y}.
\end{equation}
Equation (\ref{eq15}) is a generalization of the phase-magnetic field relation in junctions between two single band superconductors. In Josephson junctions made of multiband superconductors, there exist fractional Josephson vortices according to Eq. (\ref{eq15}), as will be discussed in Sec. XI.


Applying the Euler-Lagrangian equation with respect to $A_{b,z}$, we obtain the Ampere's law
\begin{equation}\label{eq16}
\partial _xB_{b,y}=\sin\phi_{\text{s1}}+J_{\text{s2}}\sin\phi_{\text{s2}}+\partial _tE_{b,y}.
\end{equation}
\begin{widetext}
The inter-band Josephson current $J_{12}\sin(\phi_{s1}-\phi_{s2})$ does not enter because it does not couple with the gauge field. With the help of Eqs. (\ref{eq12}) and (\ref{eq15}), we then arrive at the first equation for the gauge invariant phase difference $\phi_{si}$
\begin{equation}\label{eq17}
\frac{1}{ C_b}\left(\frac{\partial _x^2\phi _{\text{s1}}}{\zeta _1}+\frac{\partial _x^2\phi _{\text{s2}}}{\zeta _2}\right)=\sin\phi_{\text{s1}}+J_{\text{s2}}\sin\phi_{\text{s2}}+\frac{1}{ C_e}\left(\frac{\partial _t^2\phi _{\text{s1}}}{\alpha _1}+\frac{\partial _t^2\phi _{\text{s2}}}{\alpha _2}\right).
\end{equation}

We still need one more equation for the gauge invariant phase difference. This can be derived by variation of $\mathcal{L}$ with respect to $\theta_i$ and $\theta_s$, which yields
\begin{equation}\label{eq18}
-\frac{1}{\epsilon _b\alpha _s}\partial _t\left[A_0^L(r)+\partial _t\theta _s\right]-\frac{1}{\zeta _s}\partial _x\left[A_x^L-\partial _x\theta _s\right]+\sin\phi_{\text{s1}}+J_{\text{s2}}\sin\phi_{\text{s2}}=0,
\end{equation}
\begin{equation}\label{eq19}
-\frac{1}{\epsilon _b\alpha _1}\partial _t\left[A_0^R(r)+\partial _t\theta_1\right]-\frac{1}{\zeta _1}\partial _x\left[A_x^R-\partial _x\theta _1\right]-\sin\phi_{\text{s1}}-J_{12}\sin\phi_{12}=0,
\end{equation}
\begin{equation}\label{eq20}
-\frac{1}{\epsilon _b\alpha _2}\partial _t\left[A_0^R(r)+\partial _t\theta _2\right]-\frac{1}{\zeta _2}\partial _x\left[A_x^R-\partial _x\theta _2\right]-J_{\text{s2}}\sin\phi_{\text{s2}}+J_{12}\sin\phi_{12}=0.
\end{equation}
Subtracting $\zeta _1/\zeta _s\times$ Eq.(\ref{eq19}) from Eq. (\ref{eq18}), and using $A_0^B(r)+\partial _t\theta _s=-\alpha _s\epsilon _bE_b$, we obtain another equation for phases
\begin{align}\label{eq21}
\nonumber - \left( {\frac{{{\zeta _1}}}{{{\zeta _s}}}\frac{1}{{{\epsilon_b}{\alpha _1}}} + \frac{{{\zeta _1}}}{{{\zeta _s}}}\frac{{{\alpha _s}}}{{{\alpha _1}}} - 1} \right)\frac{1}{{{C_e}}}\left( {\frac{1}{{{\alpha _1}}}\partial _t^2{\phi _{{\rm{s1}}}} + \frac{1}{{{\alpha _2}}}\partial _t^2{\phi _{{\rm{s2}}}}} \right) + \frac{{{\zeta _1}}}{{{\zeta _s}}}\frac{1}{{{\epsilon_b}{\alpha _1}}}\partial _t^2{\phi _{{\rm{s1}}}}
 - \frac{1}{{{\zeta _s}}}\left( { - \frac{1}{{{C_b}}}\left( {\frac{1}{{{\zeta _1}}}\partial _x^2{\phi _{{\rm{s1}}}} + \frac{1}{{{\zeta _2}}}\partial _x^2{\phi _{{\rm{s2}}}}} \right) + \partial _x^2{\phi _{{\rm{s1}}}}} \right)\\
 + {\sin}{\phi _{{\rm{s1}}}}{\rm{ + }}{{{J}}_{{\rm{s2}}}}{\sin}{\phi _{{\rm{s2}}}}+ \frac{{{\zeta _{1}}}}{{{\zeta _{s}}}}\left( {{\sin}{\phi _{{\rm{s1}}}} + {J_{{\rm{12}}}}\sin \left( {\phi _{{\rm{s1}}}- {\phi _{{\rm{s2}}}}} \right)} \right){\rm{ = 0}}.
\end{align}
\end{widetext}
Equation (\ref{eq16}) and Eq. (\ref{eq21}) together with boundary condition (derived below) completely describe the phase dynamics in the junction. In the limit of no charge neutrality breaking $\alpha\rightarrow 0$, $\partial_t \phi_{si}=E_{b,z}$. When $|J_{12}|\gg J_{si}$, we have $\phi_{s1}=\phi_{s2}$ for $J_{12}>0$ ($s++$ pairing symmetry) and  $\phi_{s1}=\phi_{s2}+\pi$ for $J_{12}<0$ ($s\pm$ pairing symmetry). In these limits, the dynamics of the junction reduces to a single junction version
\begin{equation}\label{eq22}
\lambda_e^2\partial _x^2\phi _{\text{s1}}=J_{e}\sin\phi_{\text{s1}}+\partial _t^2\phi _{\text{s1}},
\end{equation}
with an effect Josephson coupling $J_{e}=1+\text{sign}(J_{12})J_{s2}$ and an effective penetration depth $\lambda_e=\sqrt{(\zeta_1^{-1}+\zeta_2^{-1})/C_b}$, where $\text{sign}[x]=-1$ for $x<0$ and $\text{sign}[x]=1$ for $x>0$. Here $J_{e}$ is negative for $J_{s2}<J_{s1}$ for $s\pm$ pairing symmetry, and Eq. (\ref{eq22}) describes the phase dynamics in a $\pi$ junction\cite{Chen09,Parker09,Linder09,Chen11}. However, if we express the dynamical equation in terms of $\phi_{s2}$, the sign of the Josephson current becomes positive and we have a conventional Josephson junction.

Dissipations can be introduced through a dissipation function
\begin{equation}\label{eq22a}
\mathcal{D}=\frac{1}{2}\beta_d  E_b^2,
\end{equation}
where $\beta_d$ is a damping coefficient. The equation of motion in the presence of dissipation can be derived similarly, using the Euler-Lagrange equation with dissipation
\begin{equation}\label{eq22b}
\frac{\delta \mathcal{L}}{\delta  \theta_i }-\frac{\partial }{\partial x}\left[\frac{\delta \mathcal{L}}{\delta \left(\partial _{x }\theta_i \right)}\right]-\frac{\partial }{\partial t}\left[\frac{\delta \mathcal{L}}{\delta \left(\partial _{t }\theta_i \right)}\right]=\frac{\delta \mathcal{D}}{\delta \left(\partial _t\theta_i \right)}.
\end{equation}

\section{Boundary condition}
Boundary conditions are crucial to determine the dynamics inside the junction. For a conventional single band junction, the boundary condition is given by $\partial_x \phi=2\pi B_a(2\lambda+b)/\Phi_0$ because the radiation effect is weak\cite{Bulaevskii06}, where $B_a$ is the applied field, and $\lambda$ is the penetration depth. The boundary condition for the multiband junctions cannot be generalized straightforwardly from single band cases. Using the Usadel equation, the boundary condition for multiband superconductors is developed in Ref.\cite{Brinkman04}. In the present work, we assume that the supercurrent for each band vanishes at the left and right boundary of the two-band superconductor, as shown in Fig. \ref{f1}, and we have
\begin{equation}\label{eq23}
 \left({A_x^T}-\frac{\Phi _0}{2\pi  }\partial_x \theta _1\right)=\left({A_x^T}-\frac{\Phi _0}{2\pi  }\partial_x \theta _2\right)=0,
\end{equation}
which yields $\partial_x\phi_{s1}=\partial_x\phi_{s2}$ at the boundary using Eq. (\ref{eq8}). Then from Eq. (\ref{eq15}), we derive the boundary condition
\begin{equation}\label{eq24}
\partial_x\phi_{s1}=\partial_x\phi_{s2}=\frac{C_b}{\zeta_1^{-1}+\zeta_2^{-1}} \left(B_a\pm L J_{\text{ext}}/2\right),
\end{equation}
where $J_{\text{ext}}$ is the bias current and $L$ is the length of the junction. The second term in the parenthesis at the right hand side of Eq. (\ref{eq24}) accounts for the magnetic field induced by the bias current. The magnetic field inside the junction $B_{b,y}$ is different from the applied magnetic field $B_a$ due to the screening by Josephson current.

\begin{figure}[b]
\psfig{figure=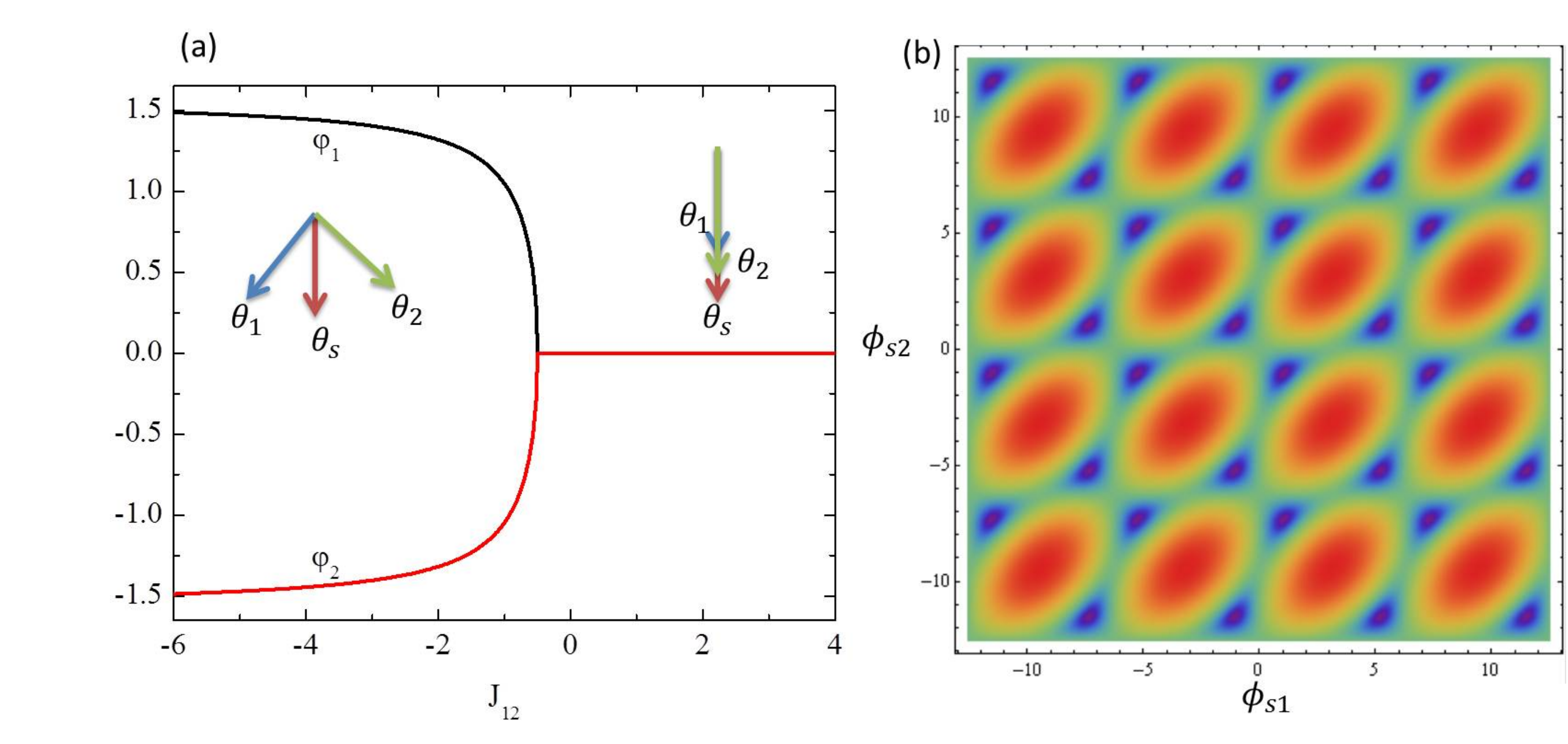,width=\columnwidth} \caption{\label{f2} (color online). (a) Phase diagram of Josephson junctions with $s\pm$ pairing symmetry. Here $J_{s2}=J_{s1}=1$. For $J_{12}<-0.5$, TRSB occurs and the transition is continuous. There are two degenerate ground states $\hat{\phi}=(\phi_{s1}, \phi_{s2})$ and $-\hat{\phi}$. Only one configuration of phases in the ground state with TRSB is shown in the figure. (b) Energy landscape in the TRSB state with $J_{12}=-1$ and $J_{s2}=1$. There are many degenerate and separated energy minima (blue region), $\hat{\phi}$, $-\hat{\phi}$ and other minima obtained by shifting phases by $2n\pi$ with an integer $n$. To contrast the minima, we plot $\log(E+1.6)$.}
\end{figure} 

\section{Ground state}
In the absence of applied magnetic fields $B_a=0$ and bias current $J_{\text{ext}}=0$, the ground state is determined by minimizing the Josephson energy
\begin{equation}\label{eq25}
E=-\cos\phi_{\text{s1}}-J_{\text{s2}}\cos\phi_{\text{s2}}-J_{12}\cos(\phi_{s1}-\phi_{s2}),
\end{equation}
which yields
\begin{equation}\label{eq26}
\sin\phi _{\text{s1}}+J_{\text{s2}}\sin\phi_{\text{s2}}=0, 
\end{equation}
\begin{equation}\label{eq27}
 \sin\phi_{\text{s1}}+J_{12}\sin \left(\phi  _{\text{s1}}-\phi  _{\text{s2}}\right)=0.
\end{equation}
We see that if $\hat{\phi}=(\phi_{s1}, \phi_{s2})$ is a solution, then $\hat{\phi}_2=-\hat{\phi}=(-\phi_{s1}, -\phi_{s2})$ is also a solution. If these two solutions are distinct, which means that one cannot obtain $\hat{\phi}_2$ from $\hat{\phi}$ by changing the phases by $2n\pi$ with an integer $n$, then TRS is broken in the junction.

In one TRSB state $(\phi_{s1}, \phi_{s2})$, the inter-junction Josephson current flows from the second band in the two-band superconductor to the $s$-wave superconductor, and then flows back to the first band in the two-band superconductor. The net current is finally balanced by the interband Josephson current from the first band to the second band in the two-band superconductor. In the other TRSB state $(-\phi_{s1}, -\phi_{s2})$, the supercurrent flow reverses direction. The meaning of TRSB becomes explicit in this picture. Please note that the supercurrent loop flows in the band space, thus is not coupled with the gauge field. No spontaneous magnetic flux is induced for spatially homogeneous systems. However as will be demonstrated below, in the ground state with TRSB, spontaneous magnetic flux will be induced when the Josephson coupling is perturbed locally.

For $J_{12}>0$, the ground state is trivial $\phi_{s1}=\phi_{s2}=0$. However for $J_{12}<0$, non-trivial solution for $\phi_{si}$ occurs due to the frustrated interaction. Let us consider $J_{12}<0$ and $|J_{12}|\gg J_{si}$. In this case $\phi _{\text{s1}}=\phi _{\text{s2}}+\pi +\delta$ with $\delta\ll 1$. We then expand Eqs. (\ref{eq26}) and (\ref{eq27}) in terms of $\delta$, and obtain 
\begin{equation}\label{eq28}
\cos\phi_{\text{s2}}=\left(1-J_{\text{s2}}\right)J_{12}/J_{\text{s2}},
\end{equation}
when $J_{12}>-\left|{J_{\text{s2}}}/({1-J_{\text{s2}}})\right|$. Otherwise $\phi_{s2}=0$ if $J_{s2}>J_{s1}$ and  $\phi_{s2}=\pi$ if $J_{s2}<J_{s1}$. Thus TRSB occurs at $J_{12}=-\left|{J_{\text{s2}}}/({1-J_{\text{s2}}})\right|$. The condition $J_{12}\gg J_{si}$ is satisfied when $J_{s1}$ and $J_{s2}$ are comparable. This is reasonable since frustration is maximized when $J_{s1}\sim J_{s2}$.

For a symmetric coupling $J_{s2}=J_{s1}=1$, we have $\phi_{s1}=-\phi_{s2}$ and the ground state can be found exactly. For $J_{12}>{-0.5}$, we have TRS state with $\phi_{s1}=\phi_{s2}=0$. For $J_{12}\leq-0.5$, TRSB occurs with $\cos\phi_{s1}=-{1}/({2 J_{12}})$. A typical phase diagram and the the energy landscape are shown in Fig. \ref{f2}.

\section{Critical current}
\begin{figure}[b]
\psfig{figure=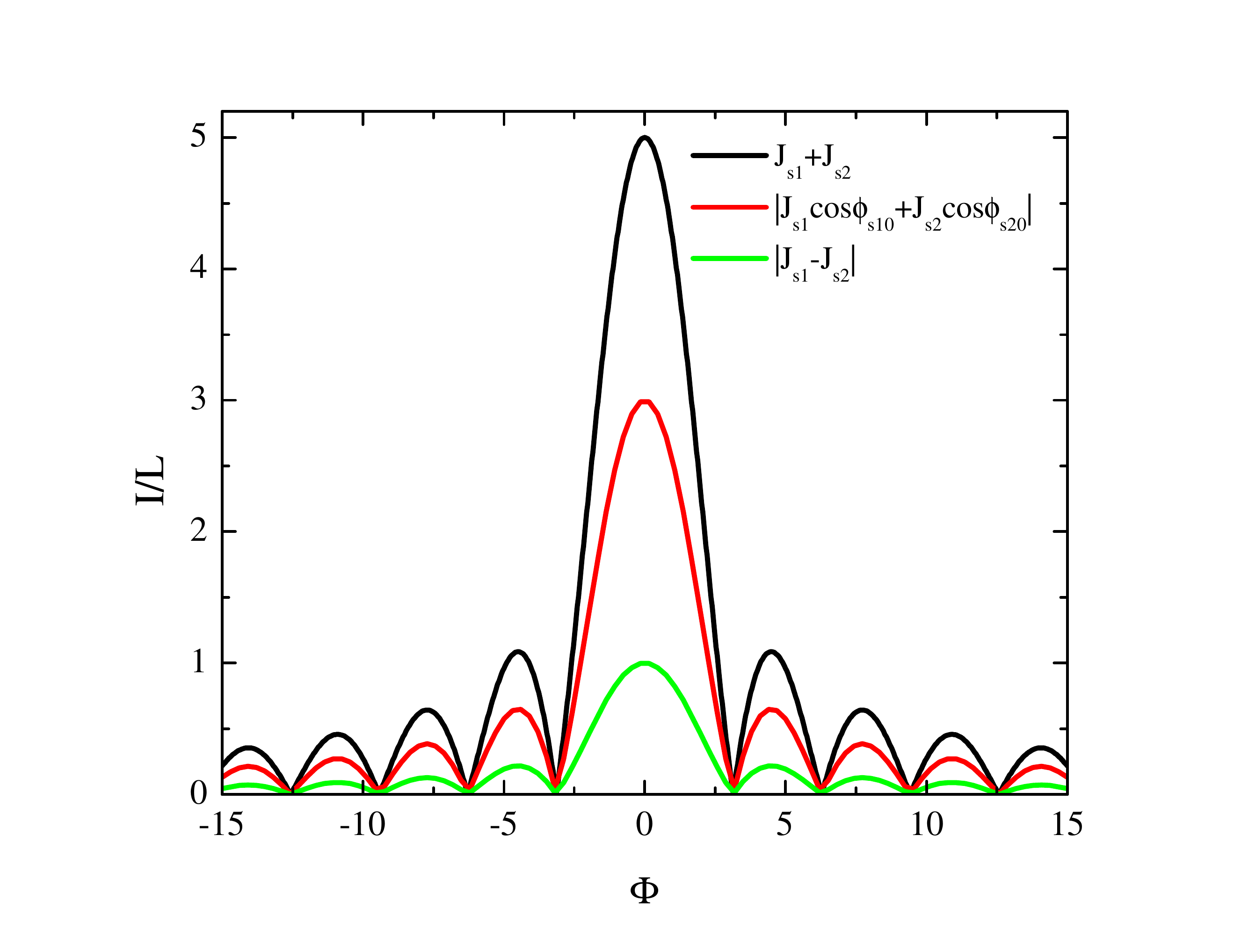,width=\columnwidth} \caption{\label{f3} (color online). Dependence of critical current on magnetic fields in three different regimes: $J_{12}>0$ with $\phi_{s1}=\phi_{s2}$, $-J_{12}\gg J_{si}$ with $\phi_{s1}=\phi_{s2}+\pi$ and the TRSB state.}
\end{figure}

In this section, we calculate the dependence of the critical current on the applied magnetic fields. Because of the interference between two tunnelling channels $J_{s1}\sin\phi_{s1}$ and $J_{s2}\sin\phi_{s2}$, the critical current of the junction depends on the phase difference between two tunnelling channels. We consider a short junction $L$ where the screening current can be neglected. [Precisely, $L$ should be much smaller than the longer length in Eqs. (\ref{eq51}) and (\ref{eq52})]. The presence of magnetic field induces a spatial variation of phases. Without the screening effect, the phase can be written as
\begin{equation}\label{eq30}
\phi _{\text{s1}}=\phi _{\text{s10}}+k_Bx+\phi _I; \text{and  }\phi _{\text{s2}}=\phi _{\text{s20}}+k_Bx+\phi _I,
\end{equation}
where $k_B=C_b B_a/\left({\zeta _1}^{-1}+{\zeta _2}^{-1}\right)$  is the phase gradient created by the applied field according to the boundary condition Eq. (\ref{eq24}), $\phi_I$ is the phase created by external current and $\phi_{si0}$ is the phase in the ground state. We have neglected the magnetic field induced by the external current because it is weak compared to $B_a$. The critical current $I$ is then expressed as
\begin{align}\label{eq31}
\nonumber\frac{I}{L} = \frac{1}{L}\int_0^L {\left( {\sin {\phi _{{\rm{s1}}}} + {J_{{\rm{s2}}}}\sin {\phi _{{\rm{s2}}}}} \right)} dx\\
 =\left| \frac{{\sin \Phi }}{\Phi } \left[{\cos  {{\phi _{{\rm{s10}}}}}  + {J_{{\rm{s2}}}}\cos  \phi _{{\rm{s20}}}}\right] \right|,
\end{align}
where we have used the ground state condition Eqs. (\ref{eq26}) and (\ref{eq27}), and $\Phi\equiv k_BL/2$. Besides the conventional factor responsible for the Fraunhofer pattern, we have additional term accounting for the interference between different tunnelling channels. When $J_{12}>0$ for $s++$ superconductors, the two channels add constructively, while for $s\pm$ superconductors with $-J_{12}\gg J_{si}$, we have $\phi_{s1}\approx\pi+\phi_{s2}$, and the two channels cancel destructively. The TRSB state with finite $0<\phi_{si0}<\pi$ interpolates these two limiting cases. We also note that the two different TRSB states $\hat{\phi}$ and $-\hat{\phi}$ has the same critical current. The dependence of critical current on magnetic field is depicted in Fig. \ref{f3} for three different cases.

\section{Shapiro steps}
In this section, we investigate the response of the junction to external microwave irradiation. When the external irradiation is locked with the internal plasma oscillation, current steps are induced known as the Shapiro steps\cite{Shapiro63}. In the presence of incident waves with frequency $\omega$, the voltage across the junction can be written as $E=E_{dc}+\tilde{E}\sin(\omega t)$. According to the generalized ac Josephson relation, the gauge-invariant phase differences can be written as
\begin{equation}\label{eq61}
\phi _{\text{s1}=}\phi _{\text{s10}}+\omega _0t+A \sin (\text{$\omega $t})+\phi _I,
\end{equation}
\begin{equation}\label{eq62}
\phi _{\text{s2}}=\phi _{\text{s20}}+\omega _0t+A \sin (\text{$\omega $t})+\phi _I,
\end{equation}
with $\omega _0={C_e}(\alpha _1^{-1}+\alpha _2^{-1})^{-1}E_{\text{dc}}$ and $A={C_e}(\alpha _1^{-1}+\alpha _2^{-1})^{-1}{\tilde{E}}/{\omega }$, and $\phi_I$ an arbitrarily relative phase between the Josephson oscillation and incident wave. Experimentally, when one fixes the voltage and tunes the current, $\phi_I$ will adjust correspondingly. The Josephson current then is given by 
\begin{align}\label{eq63}
\nonumber  I =\left\langle \left(\sin\phi_{\text{s1}}+J_{\text{s2}}\sin\phi_{\text{s2}}\right)\right\rangle _t=\\
\sum _n\left\{\cos \left(\phi _{\text{s10}}\right)+J_{\text{s2}}\cos \left(\phi _{\text{s20}}\right)\right\}J_n(A)\sin \left[\left(\omega _0-n \omega \right)t+\phi _I \right]
\end{align}
where $J_n$ is the Bessel function of the first kind and we have again used the ground state condition Eqs. (\ref{eq26}) and (\ref{eq27}). $\left\langle\cdots\right\rangle_t$ denotes time average. When the resonance condition
\begin{equation}\label{eq64a}
{C_e}\left(\alpha _1^{-1}+\alpha _2^{-1}\right)^{-1}E_{\text{dc}}=n \omega,
\end{equation}
is satisfied, the Shapiro steps appear with the height
\begin{equation}\label{eq65}
I_s=2\left|\left[\cos \left(\phi _{\text{s10}}\right)+J_{\text{s2}}\cos \left(\phi _{\text{s20}}\right)\right]J_n(A)\right|.
\end{equation}
Again the height of the Shapiro steps depends on the relative phase of the different channels, similar to that in the critical current in Eq. (\ref{eq31}).

\begin{figure}[t]
\psfig{figure=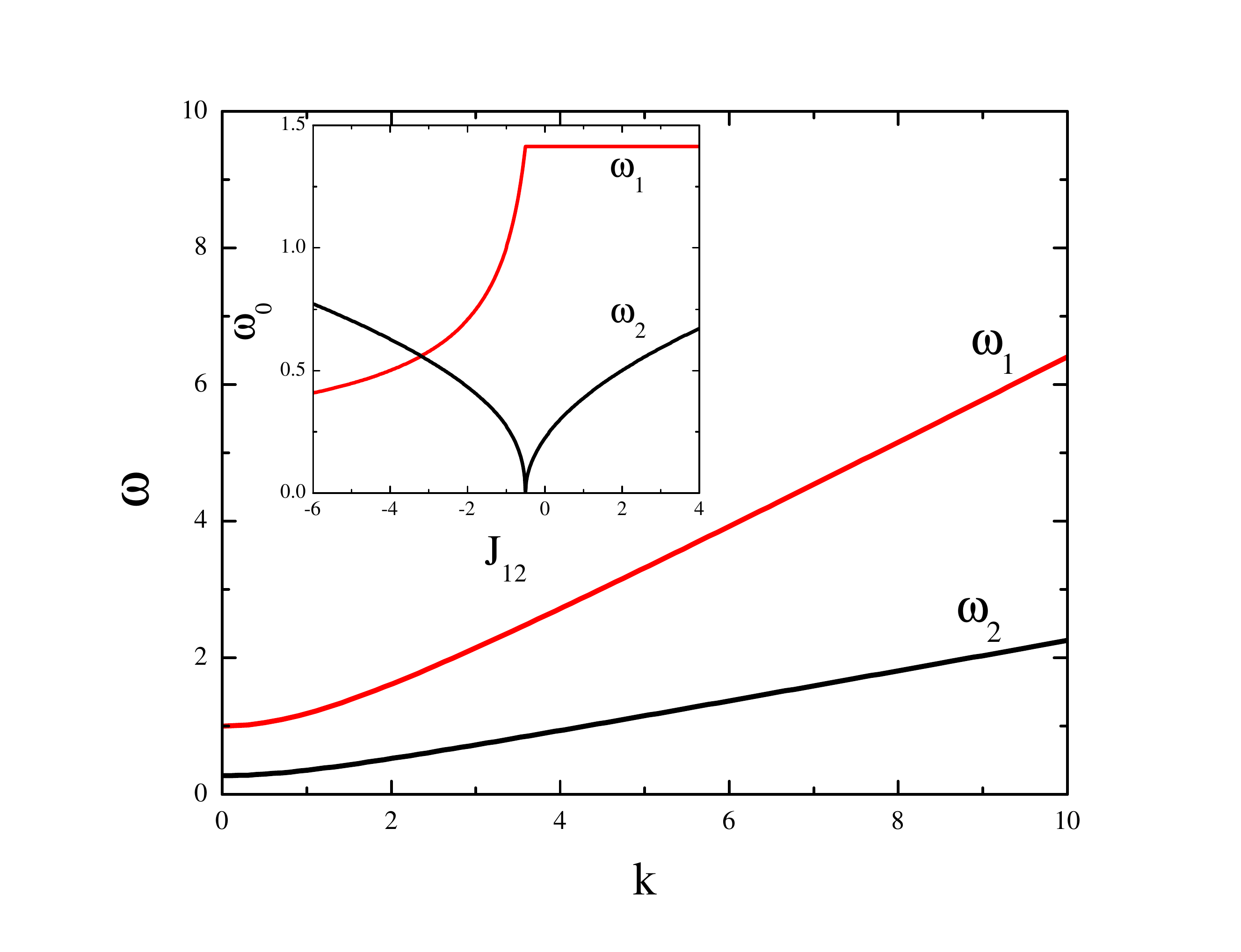,width=\columnwidth} \caption{\label{f7} (color online). Dispersion of two Josephson plasma modes according to Eq. (\ref{eq49}) and Eq. (\ref{eq50}). We use $\zeta=1$, $\alpha=0.05$ and $\epsilon_b=1$ in the plot. Inset is the energy gap $\omega_1(k=0)$ and $\omega_2(k=0)$ as a function of $J_{12}$. The branch $\omega_2(k)$ becomes massless at the TRSB transition.}
\end{figure} 

\section{Josephson Plasma mode}
The low energy collective oscillation in a Josephson junction is called Josephson plasma, which is composite waves of electromagnetic fields and Josephson current. It determines lower-energy physical properties and transport properties of the junction, such as the Fiske and Eck resonances\cite{Cirillo98}. Moreover, the dispersion for the plasma modes can be measured by Josephson plasma resonance\cite{Dahm68}, which provides a useful tool to extract physical parameters of the junction. In this section, we calculate the dispersion of the Josephson plasma in the heterotic junction between a multiband superconductor and a single band superconductor.

For convenience of calculations, we take $\alpha_s=\alpha_1=\alpha_2=\alpha\ll 1$, $\zeta_s=\zeta_1=\zeta_2=\zeta$ and $J_{s1}=J_{s2}=1$. We expand the phases around their ground state value $\bar\phi_{si}$, $\phi_{si}=\bar\phi_{si}+\varphi_{si}$, with $\bar\phi_{s2}=-\bar\phi_{s1}=\phi_0$. We then take the Fourier transform of $\varphi_{si}$, $\varphi_{si}(x,t)\sim \varphi_{si}(k,\omega)\exp[i(kx-\omega t)]$. Substituting into Eqs. (\ref{eq17}) and (\ref{eq21}), we have the following equations 
\begin{equation}\label{eq47}
\left(\frac{-k^2}{ 3\zeta+2}+\frac{\omega ^2}{ 2}-\cos\phi_0\right) \left(\varphi _{\text{s1}}+\varphi _{\text{s2}}\right)=0,
\end{equation}
\begin{align}\label{eq48}
\nonumber\left( {\frac{{ - 1}}{{2{\epsilon_b}\alpha }}{\omega ^2} + \frac{{{k^2}}}{\zeta }\frac{{3\zeta  + 1}}{{3\zeta  + 2}} + 2\cos {\phi _0} + {J_{12}}\cos (2{\phi _0})} \right){\varphi _{s1}}+\\
\left( {\frac{1}{{2{\epsilon_b}\alpha }}{\omega ^2} - \frac{{{k^2}}}{\zeta }\frac{1}{{3\zeta  + 2}} + \cos {\phi _0} - {J_{12}}\cos (2{\phi _0})} \right){\varphi _{s2}} = 0
\end{align}
We then derive two branches of the dispersion relation
\begin{equation}\label{eq49}
\omega_1 ^2=2\cos \theta _0+\frac{2k^2}{3\zeta +2},
\end{equation}
\begin{equation}\label{eq50}
\omega_2 ^2={\epsilon _b\alpha }\left[\frac{k^2}{\zeta }+\cos \phi _0+2J_{12}\cos (2\phi _0)\right],
\end{equation}
with $\cos\phi_0=-1/(2J_{12})$. The results are displayed in Fig. \ref{f7}. There are two Swihart velocities with $c_1=\sqrt{2/(3+2\zeta)}$ and $c_2=\sqrt{\epsilon_b\alpha/\zeta}$. The splitting of the Josephson plasma into two branches thus supports Cherenkov radiation in the junction, similar to that in a stack of Josephson junctions\cite{Mints95,Hechtfischer97}. In the TRSB state, the plasma modes for the two degenerate TRSB pair states $\hat{\phi}$ and $-\hat{\phi}$ are identical. The mode $\omega_1$ corresponds to the in-phase oscillation of $\varphi_{s1}$ and $\varphi_{s2}$, and is always massive. The mode $\omega_2$ corresponds to the out-of-phase oscillation, which is similar to the Leggett mode\cite{Leggett66} in bulk multiband superconductors. In the limit of $\alpha\rightarrow 0$, this mode disappears because the out-of-phase oscillation is forbidden according to the ac Josephson relation $\partial_t\phi_{s1}=\partial_t\phi_{s2}=E_{b,z}$. In the presence of charge neutrality breaking, the mode $\omega_2$ becomes massless at the TRSB transition $\cos\phi_0+2J_{12}\cos(2\phi_0)=0$. Close to the TRSB transition, the angle between $\phi_{s1}$ and $\phi_{s2}$ is about to grow continuously from $0$ to a finite value as shown in Fig. \ref{f2}(a), thus the out-of-phase oscillation becomes a soft mode. 

Near a continuous phase transition, the energy cost for the excitation of perturbations is close to zero, thus can be easily excited, which in turn tends to destroy the ordered state. At the second-order TRSB transition, the out-of-phase oscillation $\phi_{s1}$ and $\phi_{s2}$ does not cost energy and the phase rigidity for the out-of-phase oscillation vanishes. (The phase rigidity for the in-phase oscillation is nonzero, see Fig. \ref{f7}.) In bulk superconductors, this phase rigidity is proportional to the superfluid density and the mass of photon according to the Anderson-Higgs mechanism\cite{SimonsQFT}. The vanish of the phase rigidity in the junction means that the photon mass becomes zero.  

The massless plasma mode at the TRSB transition in Eq. (\ref{eq50}) is remarkable. In a conventional Josephson junction where the superconductivity is suppressed in the barrier, the mass of the Josephson plasma is reduced compared to that in bulk, but the plasma excitation still has an energy gap proportional to the Josephson coupling. Here we show for the first time that there exists genuine massless photon in a junction with TRSB. The massless plasma mode is analogous to the massless Leggett boson discussed in bulk superconductors with TRSB.\cite{szlin11b}. The difference is that the Leggett mode is a purely phase mode, while the plasma mode is a composite mode of superconducting phase and gauge fields. The massless Josephson plasma mode gives rise novel phenomena as will be elucidated in the next section.

\section{Josephson penetration depth and $H_{c1}$} 
In this section, we calculate the Josephson penetration depth and the threshold magnetic field $H_{c1}$ where magnetic flux starts to penetrate into the junctions\cite{TinkhamBook}. The penetration depth is determined by the mass of the photon in the junction according to the London equation. By setting $\omega=0$ in Eq. (\ref{eq49}) and Eq. (\ref{eq50}), we obtain the Josephson penetration depth
\begin{equation}\label{eq51}
\lambda _{\text{J1}}=1/\sqrt{(3\zeta +2)\cos\phi_0 }.
\end{equation}
\begin{equation}\label{eq52}
\lambda _{\text{J2}}=1/\sqrt{\zeta \left[\cos\phi_0+2J_{12}\cos (2\phi _0)\right]}.
\end{equation}
The Josephson penetration depth of the junction is given by the larger value of $\lambda _{\text{J1}}$ and $\lambda _{\text{J2}}$. Being a static property, the Josephson penetration depth is independent of $\alpha$. At the TRSB transition, $\lambda_{J2}$ diverges as a consequence of the vanishing photon mass. 

In conventional type II superconductors, magnetic fields can penetrate into the superconductor only when they are larger than a critical value $H_{c1}$, due to the screening of fields as a consequence of massive photons. This becomes evident when we look at $H_{c1}\sim \Phi_0/\lambda_L^2$ obtained under the London approximation. Here the London penetration depth $\lambda_L$ is inverse proportional to the mass of photon $\lambda_L\sim 1/\sqrt{m_p}$. For a conventional Josephson junction, $H_{c1}$ is reduced as a result of weakened superconductivity in the junction, but $H_{c1}$ is nonzero $H_{c1}>0$. In junctions with massless photons at TRSB transition, the magnetic field can penetrate easily into the system, and the lower critical field becomes zero, $H_{c1}=0$.

\section{McCumber solution}
In the absence of applied magnetic fields, the phases are uniform along the $x$ direction for a homogeneous junction. When the external current exceeds the critical current $I_c$ of the junction Eq. (\ref{eq30}), the junction switches into a resistive state and the phases start to rotate, and such state is called the McCumber state\cite{Kleiner00}. For an overdamped junction ($\beta_d$ in Eq. (\ref{eq22a}) is large $\beta_d\gg 1$. The standard McCumber number is given by $1/\beta_d^2$.), the junction returns to zero-voltage state when the external current is smaller than $I_c$. However, for an underdamped junction with $\beta_d\ll 1$, the system remains resistive, until a current much smaller than $I_c$ where the system transits into zero-voltage state again\cite{Hu10}.  In this section, we calculate the $\emph{IV}$ curve in the McCumber state. 

The phases in the McCumber state, for a symmetric coupling $J_{s2}=J_{s1}$, can be written as 
\begin{equation}\label{eq52a}
\phi _{\text{s1}}=\phi _0+\omega  t+\text{Re}\left[-i \varphi _{\text{s1}}\exp (i \omega  t)\right],
\end{equation}
\begin{equation}\label{eq52b}
\phi _{\text{s2}}=-\phi_0+\omega  t+\text{Re}\left[-i \varphi _{\text{s2}}\exp (i \omega  t)\right].
\end{equation}
Substituting Eqs. (\ref{eq52a}) and (\ref{eq52b}) into Eqs. (\ref{eq17}) and (\ref{eq21}) and approximating $\sin(\phi_{s1})\approx -i\exp[i(\phi_0+\omega t)]$, we have
\begin{align}\label{eq52c}
\nonumber\left[\frac{-\omega ^2}{2\epsilon _b\alpha }+J_{12}\cos (2\phi _0)\right]\varphi _{s1}
+\left[\frac{\omega ^2}{2\epsilon _b\alpha }-J_{12}\cos (2\phi _0)\right]\varphi _{s2}\\=-\left[2\exp({i \phi _0})+\exp({-i \phi _0})\right],
\end{align}
\begin{equation}\label{eq52d}
{\omega ^2}\left(\varphi _{\text{s1}}+\varphi _{\text{s2}}\right)=4 \cos \left(\phi _0\right).
\end{equation}
Equation (\ref{eq52c}) and (\ref{eq52d}) can be solved and we have
\begin{equation}\label{eq52e}
\varphi_{\text{s1}}=\frac{2 \cos\phi_0}{\omega^2}+\frac{2\exp({i \phi _0})+\exp({-i \phi _0})}{\omega^2/(\epsilon _b\alpha)-2 J_{12}\cos (2\phi _0)},
\end{equation}
\begin{equation}\label{eq52f}
\varphi_{\text{s2}}=\frac{2 \cos\phi_0}{\omega^2}-\frac{2\exp({i \phi _0})+\exp({-i \phi _0})}{\omega^2/(\epsilon _b\alpha)-2 J_{12}\cos (2\phi _0)}.
\end{equation}
Then the supercurrent contributed from the phase oscillation is given by
\begin{align}\label{eq52g}
\nonumber J_s=\left\langle \sin \left(\phi _{\text{s1}}\right)+J_{\text{s2}}\sin \left(\phi _{\text{s2}}\right)\right\rangle _t\\
=-\text{Re}\left[\sin\phi_0\frac{2\exp({i \phi _0})+\exp({-i \phi _0})}{\omega^2/(\epsilon _b\alpha)-2 J_{12}\cos (2\phi _0)}\right].
\end{align}
The supercurrent increases when voltage $\omega$ decreases. The total current density is $J=\beta_d\omega+J_s$, with an ohmic contribution $\beta_d\omega$.

\section{Flux flow region: Fiske and Eck resonances}
When strong magnetic fields are applied parallel to the junction, magnetic fields penetrate into the junction and form Josephson vortices. Under a transport current, the vortices are driven by the Lorentz force and move along the junction, which induces finite voltage across the junction. Then the Josephson plasma is excited. When the plasma is resonant with the vortex motion, it induces a large dc part of supercurrent, which manifests as a current step known as the Eck step\cite{Eck64}. For a junction of finite length, the plasma can be resonant with the cavity formed by the junction itself under appropriate conditions, which also induces current steps known as the Fiske steps\cite{Fiske64}. In this section, we calculate the resonances in the flux flow region. Studies of the flux-flow dynamics in two-band junctions with an emphasis on the inter-band Josephson coupling were presented in Ref. \cite{Kim12} very recently.  

\begin{figure}[t]
\psfig{figure=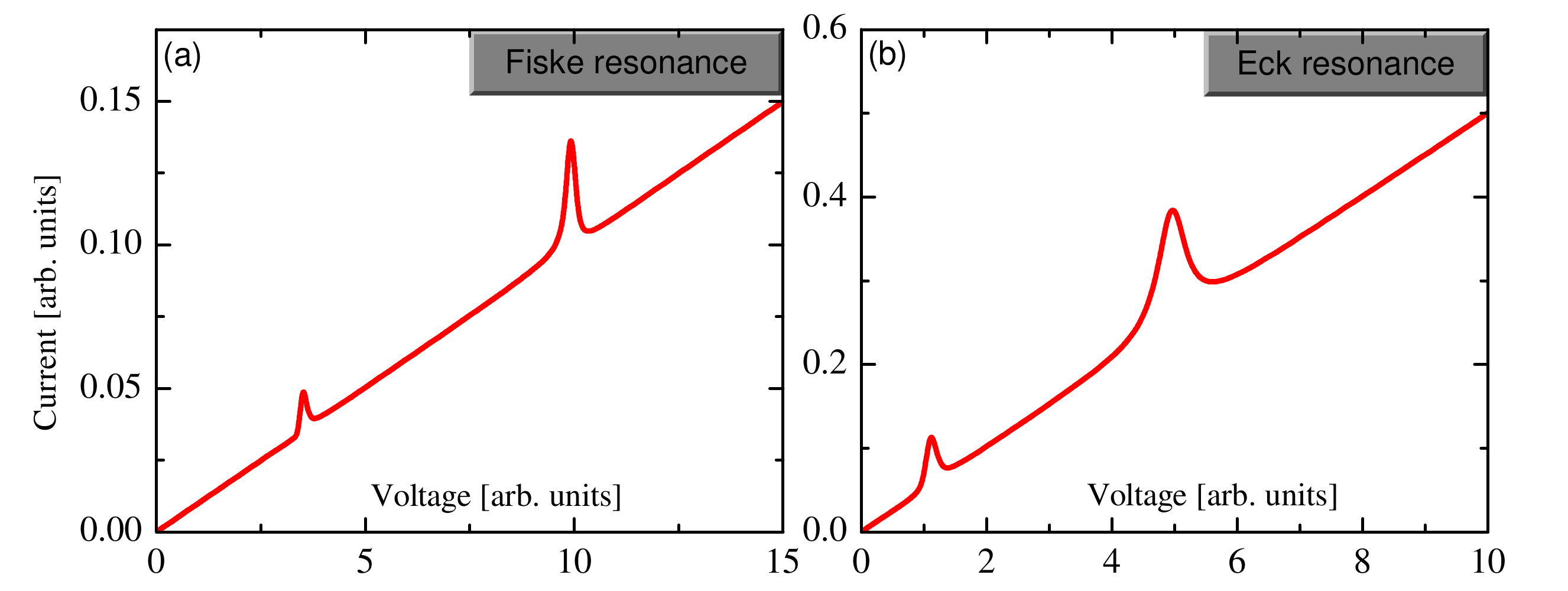,width=\columnwidth} \caption{\label{f8} (color online). (a) \emph{IV} curve in the flux flow region. The resonance peaks shown in the figure are caused by the excitation of cavity modes (Fiske resonance). Here we only plot the resonance mode $k_m=\pi/L$ with $L=0.2$. We use $\zeta=1$, $\alpha=0.05$, $\epsilon_b=1$ and $B_a=\pi$. The ohmic contribution in the figure is $\beta_d\omega$ with  $\beta_d=0.01$. (b) Eck resonance in the flux low region. We use $\zeta=1$, $\alpha=0.05$, $\epsilon_b=1$, $B_a=\pi$ and $\beta_d=0.05$. }
\end{figure}

For a junction with a finite length $L$ and with a symmetric coupling $J_{s2}=J_{s1}$, the phase in the flux flow region can be written as
\begin{equation}\label{eq53}
\phi _{\text{s1}}=\phi _0+k_Bx+\omega  t+\text{Re}\left[-i \varphi _{\text{s1}}\cos (k_m  x)\exp (i \omega  t)\right],
\end{equation}
\begin{equation}\label{eq54}
\phi _{\text{s2}}=-\phi_0+k_Bx+\omega  t+\text{Re}\left[-i \varphi _{\text{s2}}\cos (k_m  x)\exp (i \omega  t)\right],
\end{equation}
where $k_B=C_b B_a/(\zeta_1^{-1}+\zeta_2^{-1})$ is the phase gradient due to the applied field $B_a$. Here $k_m=m\pi/L$ accounts for the geometric resonances, and $\phi_0$ is the ground state value. Substituting Eqs. (\ref{eq53}) and (\ref{eq54}) into Eqs. (\ref{eq17}) and (\ref{eq21}) and using $\sin(\phi_{s1})\approx -i\exp[i(\phi_0+k_B x+\omega t)]$ to the zeroth order, we have
\begin{align}\label{eq55}
\nonumber\left[\frac{\omega ^2}{2\epsilon _b\alpha }-J_{12}\cos (2\phi _0)\right]\left(\varphi _{s2}-\varphi _{s1}\right)+\frac{k_m^2}{\zeta }\frac{3\zeta +1}{3\zeta +2}\varphi _{s1}\\
-\frac{k_m^2}{\zeta }\frac{1}{3\zeta +2}\varphi _{s2}=-\left[2\exp({i \phi _0})+\exp({-i \phi _0})\right]F_c,
\end{align}
\begin{equation}\label{eq56}
\left(\frac{-k_m^2}{3\zeta +2}+\frac{\omega ^2}{2}\right)\left(\varphi _{\text{s1}}+\varphi _{\text{s2}}\right)=2 \cos \left(\phi _0\right)F_c,
\end{equation}
where we have introduced the coupling between the flux flow and plasma oscillation at the cavity modes $k_m$
\begin{equation}\label{eq57}
F_c=\frac{2}{L}\int _0^Ldx \cos \left(k_mx\right)\exp \left(i k_Bx\right).
\end{equation}
Resonances occur at 
\begin{equation}\label{eq58}
\omega ^2=\frac{2k_m^2}{3\zeta +2},
\end{equation}
\begin{equation}\label{eq59}
\frac{1}{\epsilon _b\alpha }\omega ^2=\frac{k_m^2}{\zeta }+2J_{12}\cos (2\phi _0),
\end{equation}
that is when the plasma frequency matches the cavity frequency. Thus in the presence of the charge neutrality breaking, there are two resonant peaks for a given cavity mode $m$ because there are two different dispersion branches. The supercurrent induced by the resonance is
\begin{align}\label{eq60}
\nonumber J_s=\frac{1}{L}\int \left\langle \sin \left(\phi _{\text{s1}}\right)+J_{\text{s2}}\sin \left(\phi _{\text{s2}}\right)\right\rangle _t dx\\
=\frac{-i F_c^*}{4}\left[\varphi _{\text{s1}}\exp({-i \phi _0})+J_{\text{s2}}\varphi _{\text{s2}}\exp({i \phi _0})\right],
\end{align}
where again the terms in the square bracket accounts for the interference between two tunnelling channels. Real junction inevitably involves dissipation, and the delta-peak resonance in Eqs. (\ref{eq58}) and (\ref{eq59}) is rounded by the dissipation. A formal treatment would introduce dissipation function as discussed in Eq. (\ref{eq22a}). Here we introduce phenomenologically the dissipation by the replacement $\omega ^2\leftarrow \omega ^2-i \beta_d  \omega$. We calculate the \emph{IV} curve by solving Eqs. (\ref{eq55}-\ref{eq57}) and Eq. (\ref{eq60}) numerically and the result is displayed in Fig. \ref{f8}(a), where two resonant peaks can be clearly observed.

For a long junction, the geometry resonance is not important. In this case, we need to replace the solutions in Eqs. (\ref{eq53}) and (\ref{eq54}) by $k_m\leftarrow k_B$, i.e. the plasma has the same spatial modulation as the flux flow along the $x$ direction, because the plasma oscillation is excited by the flux flow. The expressions of the resonance conditions and \emph{IV} curves are the same as those of the Fiske resonance except for the replacement of $k_m\leftarrow k_B$. When the velocity of the flux flow matches the velocity of plasma (there are two dispersion relations with two different velocities), current steps known as Eck steps appear as shown in Fig. \ref{f8}(b).

For a stack of conventional Josephson junctions, the plasma dispersion splits into $N$ branches with $N$ the number of junctions\cite{Ngai69,Sakai93,Sakai94}. It gives $N$ resonant branches in the \emph{IV} characteristic for a given mode. The experimental observation of $N$ resonant branches thus confirms the plasma splitting in junction stacks \cite{Ustinov93}. Similarly, the charge neutrality breaking effect discussed here can also be checked experimentally by measurement of doubling of resonant branches in the \emph{IV} characteristic.

\section{Topological excitations}
In the present system, the potential energy has many degenerate minima, which differs in $\phi_{si}$ by $2\pi$, as shown in Eq. (\ref{eq25}) and Fig. \ref{f2}(b). In the TRSB state, there is additional minima at $\hat{\phi}$ and $-\hat{\phi}$ due to TRSB. Thus the system supports stable topological excitations, categorized as solitons or phase kinks belonging to the homotopy class of $\pi_0(S^0)$\cite{Manton04}. There are solitons with two different types of origins, with one being the soliton between the energy minima with $2n\pi$ phase shift, and the other the soliton between two TRSB pair states, $\hat{\phi}$ and $-\hat{\phi}$. The width of the phase kink is determined by the optimization between the potential energy in Eq. (\ref{eq25}) and the energy due to the phase gradient. When the width of kink increases, the cost in the potential energy increases while the cost in energy associated with the phase gradient decreases, and vice versa. The stable kink solution compromises these two energy costs. In this section, we investigate the possible soliton solutions in the junction. For discussions of phase kink in bulk multiband superconductors, please see Ref. \cite{tanaka02,szlin11c,Vakaryuk2012}.

\begin{figure}[t]
\psfig{figure=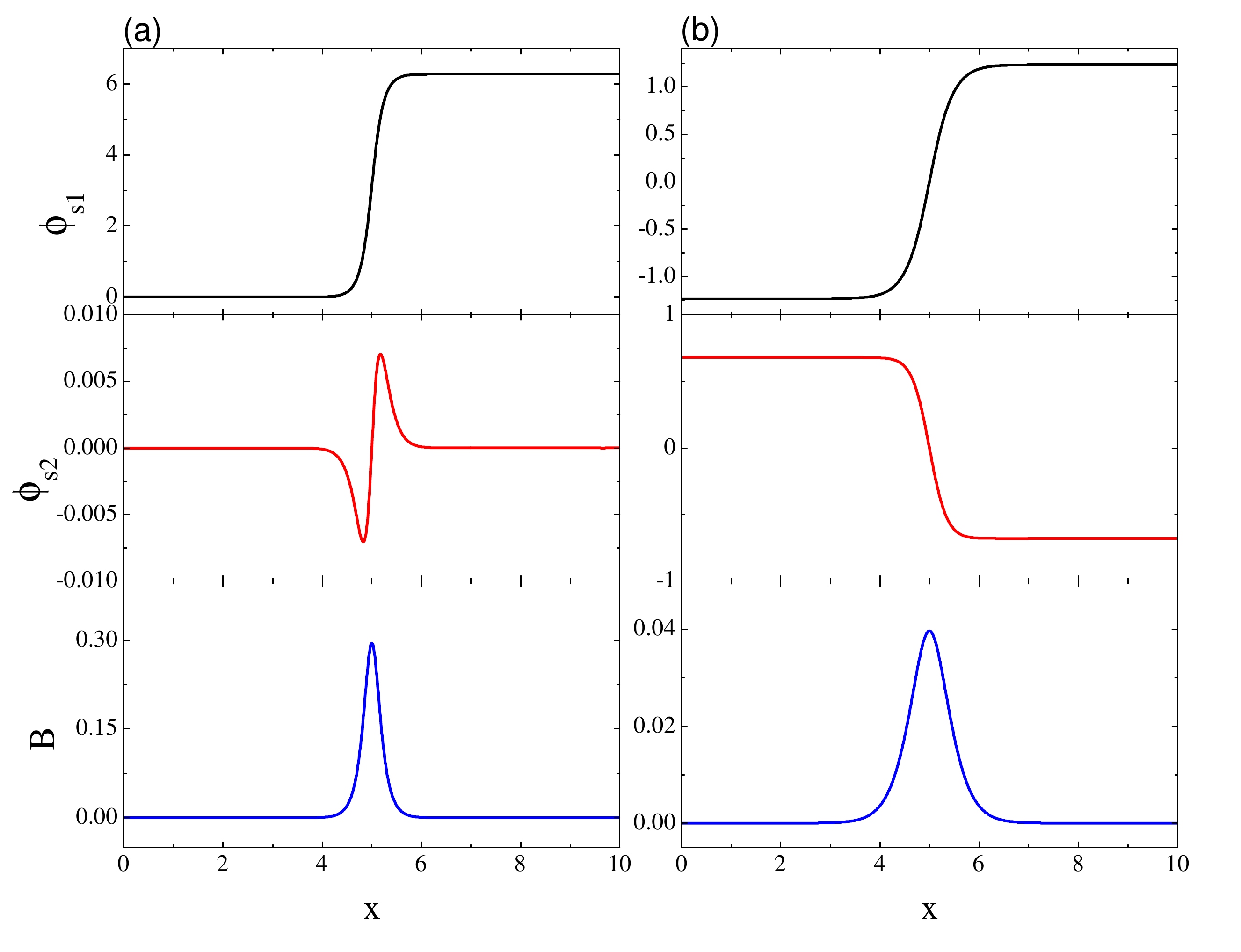,width=\columnwidth} \caption{\label{f9} (color online). (a) Profile of phases and magnetic field in (a) a conventional $2\pi$ solition, where $\phi_{\text{s1}}$ changes by $2\pi$, (b) soliton between two TRSB pair states.  Here $\zeta_1=10$, $\zeta_2=24$, $\zeta_s=20$, $J_{s1}=1.0$, $J_{s2}=1.5$. $J_{12}=1.0$ in (a) and  $J_{12}=-1.0$ in (b).}
\end{figure}

First let us consider the soliton solution by changing $\phi_{si}$ by $2n_i\pi$ with an integer $n_i$. The total magnetic flux associated with these solitons is given by
\begin{equation}\label{eq66}
\Phi(n_i)=(\zeta_1^{-1}2n_i\pi+\zeta_2^{-1}2n_2\pi)/C_b.
\end{equation}
It is fractional quantized if $n_1\neq n_2$. Unlike the fractional quantized vortices in bulk multiband superconductors\cite{Babaev02}, the energy associated with the fractional soliton in junctions is bounded, thus the excitation is thermodynamically stable. We calculate numerically a typical configuration of the soliton in the junction, where $\phi_{s1}$ changes by $2\pi$ and $\phi_{s2}$ does not change from the left edge to the right edge. The results are shown in Fig. \ref{f9}(a). The presence of solitons breaks TRS, even though in the ground state, the system has TRS.

Now let us investigate the soliton solution between two TRSB pair states $\hat{\phi}$ and $-\hat{\phi}$, when the ground breaks TRS. An exact expression can be found for the symmetric case $\zeta_s=\zeta_1=\zeta_2=\zeta$ and $J_{s1}=J_{s2}=1$. In this case $\phi_{s1}(x)=-\phi_{s2}(x)$, and the equation for the spatial variation of phase becomes
\begin{equation}\label{eq67}
-\frac{1}{\zeta }\partial _x^2\phi _{\text{s1}}+\sin\phi_{\text{s1}}+J_{12}\sin \left(2\phi  _{\text{s1}}\right)=0.
\end{equation}
We use the Bogomolny method\cite{Manton04} to find the exact solution in the following. The energy corresponding to Eq. (\ref{eq67}) is 
\begin{equation}\label{eq68}
E_s=\frac{\left(\partial _x\phi_{s1} \right)^2}{2\zeta}- (\cos\phi_{s1}-\cos\phi_0) -\frac{J_{12}}{2}[\cos (2\phi_{s1} )-\cos(2\phi_0)]
\end{equation}
where $\cos\phi_0=-1/(2J_{12})$ is the ground state value and we introduce them into $E_s$ to shift the energy minimum to $0$. We then consider the following inequality
\begin{equation}\label{eq69}
\left(\frac{1}{\sqrt{2\zeta }}\partial _x\phi_{s1} \pm \sqrt{U}\right)^2\geq 0,
\end{equation}
with $U=\left(\sqrt{-J_{12}}\cos\phi_{s1}-\frac{1}{2\sqrt{-J_{12}}}\right)^2$. Thus the energy of the kink has a lower bound
\begin{equation}\label{eq70}
E_s=\frac{1}{2\zeta}\left(\partial _x\phi_{s1} \right)^2+U\geq \pm \frac{\sqrt{2}}{\sqrt{\zeta }}\partial _x\phi_{s1} \sqrt{U}.
\end{equation}
The lower bound is reached when
\begin{equation}\label{eq71}
\partial _x\phi_{s1} =\pm\sqrt{2\zeta U}=\pm\sqrt{2\zeta }\left(\sqrt{-J_{12}}\cos\phi_{s1}-\frac{1}{2\sqrt{-J_{12}}}\right).
\end{equation}
Equation (\ref{eq71}) can be integrated explicitly, which yields a soliton solution between the two TRSB pair states
\begin{equation}\label{eq72}
\phi_{s1} =2\tan^{-1}\left[\frac{\sqrt{-1+4 J_{12}^2}}{1-2 J_{12}}\tanh\left(\pm\frac{\sqrt{\zeta }\sqrt{-1+4 J_{12}^2}}{2\sqrt{2} \sqrt{-J_{12}}}x\right)\right],
\end{equation}
where $+$ corresponds to soliton and $-$ corresponds to anti-soliton. The energy associated with the kink is 
\begin{equation}\label{eq73}
E_s=\sqrt{\frac{{2}}{{\zeta }}}\left|\left(2\sqrt{-J_{12}}\sin\phi_0-\frac{1}{\sqrt{-J_{12}}}\phi _0\right)\right|.
\end{equation}
In the symmetric case as we studied here, there is no magnetic flux associated with the soliton because the phase gradient in each channel compensates exactly $\partial_x(\phi_{s1}+\phi_{s2})=0$. However, for more general cases, soliton carries finite magnetic flux. We calculate numerically the soliton solution and the associated magnetic flux for asymmetric parameters $\zeta_1\neq\zeta_2\neq\zeta_s$, and the results are displayed in Fig. \ref{f9}(b). Suppose at the left domain, the phases belong to one ground state $-\phi_{s10}$ and $-\phi_{s20}$, and at the right domain, the phases belong to the other ground state $\phi_{s10}$ and $\phi_{s20}$, the total magnetic flux is quantized in terms of $\phi_{si0}$
 \begin{equation}\label{eq73a}
\Phi=2(\zeta_1^{-1}\phi_{s10}+\zeta_2^{-1}\phi_{s20})/C_b.
\end{equation}

The discussions of the solitons so far are restricted to the static case where the stability of solitons are guaranteed by the energy landscape of the Josephson coupling. In the dynamic region when a transport current is present, solitons can be stabilized with some voltages. Solitons can also be created due to the dynamic instability of superconducting phase in the Josephson junctions. \cite{Pagano86} The solitons start to move because of the Lorentz force. When the velocity of soliton matches the velocity of plasma, current steps known as the zero-field steps arise at these voltages\cite{Fulton73}. Due to the existence of various types of fractionally quantized solitons and the existence of two Swihart velocities for plasma in the present system, we expect appearance of many zero-field steps in the \emph{IV} curve.

The existence of solitons associated with flux in Eq. (\ref{eq66}) does not depends on the pairing symmetry of the two-band superconductors, while the soliton between two TRSB pair states in Eq. (\ref{eq72}) exists only in superconductors with $s\pm$ pairing symmetry, where TRSB in the ground state is possible.

\begin{figure}[b]
\psfig{figure=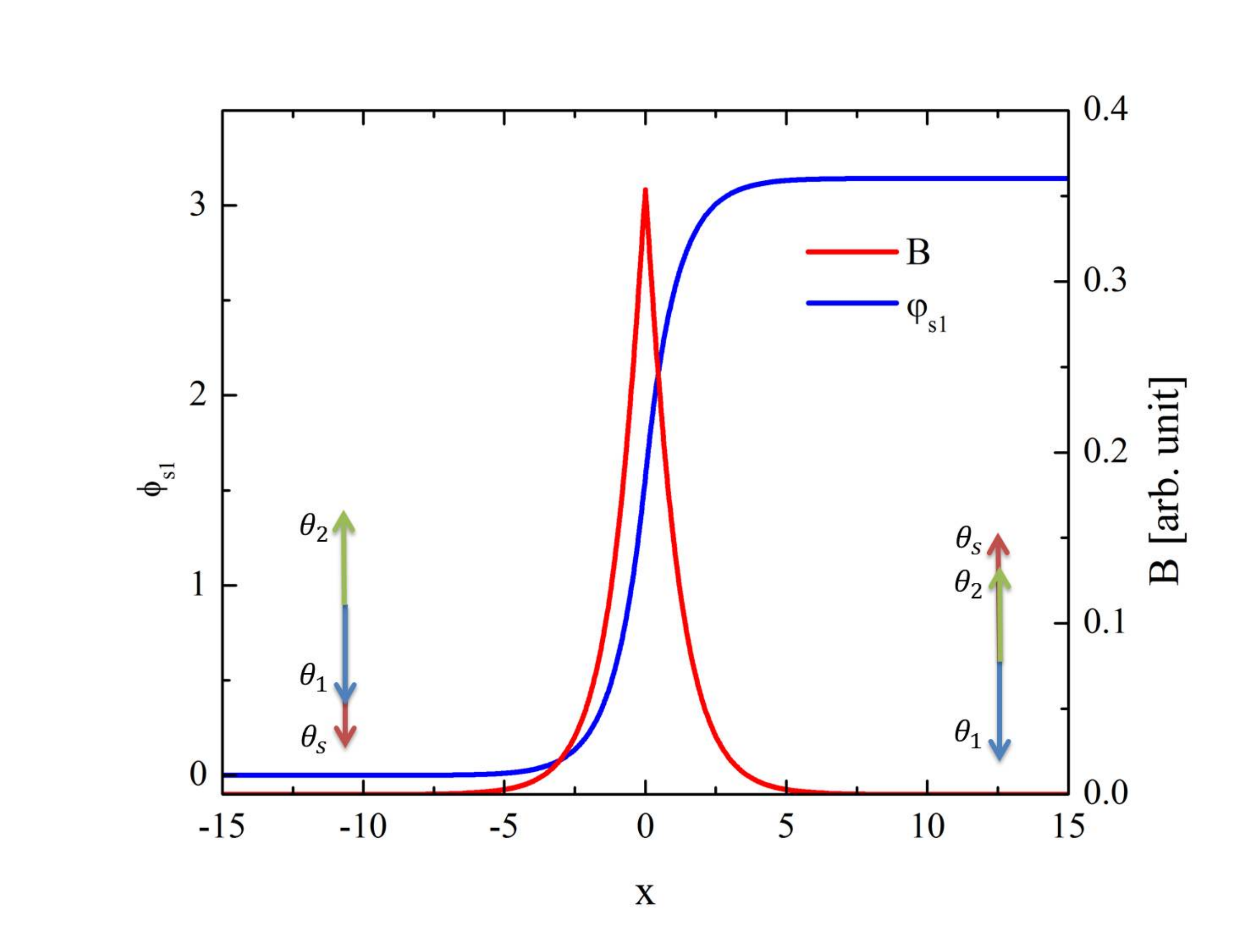,width=\columnwidth} \caption{\label{f4} (color online). Profile of the phases and magnetic field with a step modulation of the Josephson coupling in a junction with $s\pm$ pairing symmetry. Here $\phi_{s2}=\phi_{s1}+\pi$ and $\lambda_e=J'=1$.}
\end{figure}

 \section{Spontaneous magnetic flux}
In this section, we investigate the spontaneous magnetic flux in junctions unique to superconducting electrodes with the $s\pm$ pairing symmetry, which points a practically useful way to detect the pairing symmetry. We consider a junction with spatial variation of the Josephson couplings, which is likely realized in real systems due to inhomogeneity, such as variation of the thickness of barrier, or can be tuned intentionally as in Ref. \cite{Chen10}. Let us first consider a step modulation of the Josephson coupling, which is amenable to analytical calculation
\begin{equation}\label{eq32}
J_{s2}-J_{s1}=J'[2\Theta(x)-1],
\end{equation}
with the Heaviside step function $\Theta(x)$. We further assume $-J_{12}\gg J_{si}$, and the spatial variation of phase is given by according to Eq. (\ref{eq22})
\begin{equation}\label{eq33}
\lambda_e^2\partial _x^2\phi _{\text{s1}}-J'[-2\Theta(x)+1]\sin\phi_{s1}=0,
\end{equation}
with the boundary condition $\partial_x\phi_{s1}(x=\pm L/2)=0$. The length of the junction $L$ is assumed to be small $L\ll \lambda_e$. In this case, the solution can be expressed as
\begin{equation}\label{eq34}
\phi_{s1} (x)=\phi _{s10}+\eta (x) \sin\phi_{s10},
\end{equation}
with $\eta (x)<<1$. Substituting the solution into Eq. (\ref{eq33}), we obtain the equation for $\eta(x)$
\begin{equation}\label{eq35}
\lambda_e ^2\partial _x^2\eta (x)-J'(-2\Theta (x)+1) \left[1+\eta (x) \cos\phi_{s10}\right]=0.
\end{equation}
The first term in the square bracket dominates and to the first order, we can safely neglect the second term in the square bracket. Integrating along the $x$ direction yields the magnetic flux
\begin{equation}\label{eq36}
B=\frac{\zeta_1^{-1}+\zeta_2^{-1}}{C_b}\sin\phi_{s10}\partial _x\eta (x)={J'}\sin\phi_{s10}\left(-\frac{L}{2}+|x|\right).
\end{equation}
$\phi_{s10}$ is given by the condition that spatial average of Eq. (\ref{eq35}) vanishes because of the boundary condition $\partial_x\phi_{s1}=0$ as no external magnetic field is applied
\begin{equation}\label{eq36a}
\int_{-L/2}^{L/2} J'[-2\Theta (x)+1] \eta (x) \cos\phi_{s10}dx=0,
\end{equation}
which yields $\phi_{s10}=\pi/2$.

The magnetic flux inside depends on the length of junction and is not quantized for a short junction $L\ll \lambda_e$. However for a long junction $L\gg \lambda_e$, magnetic flux only occurs in the region where the Josephson coupling changes and it is quantized due to the complete screening. We investigate the magnetic flux in a long junction $L\gg\lambda_e$, where the nonlinear effect of Josephson current must be treated self-consistently. The solution can be constructed as follows: for $x<0$, 
\begin{equation}\label{eq37}
\phi _{s1}^L=4 \arctan \left[\exp \left(\frac{\sqrt{J'}}{\lambda_e }\left(x+x_0\right)\right)\right],
\end{equation}
and for $x>0$
\begin{equation}\label{eq38}
\phi _{s1}^R=4 \arctan \left[\exp \left(\frac{\sqrt{J'}}{\lambda_e }\left(x-x_0\right)\right)\right]-\pi.
\end{equation}
At $x=0$, the derivative of $\phi_{s1}$ is automatically continuous under this construction. $x_0$ is determined by the continuity condition $\phi_{s1}^L(x=0)=\phi_{s1}^R(x=0)$, which yields $x_0={\lambda_e }\ln \left(-1+\sqrt{2}\right)/\sqrt{J'}$. The total magnetic flux inside the junction is given by 
\begin{equation}\label{eq39}
\Phi=\int_{-L/2}^{L/2} B dx=\frac{\zeta_1^{-1}+\zeta_2^{-1}}{C_b}\int_0^\pi\partial_x\phi_{s1}=\frac{\zeta_1^{-1}+\zeta_2^{-1}}{C_b}\pi,
\end{equation}
is quantized since $\phi_{s1}$ run from $0$ at $x=-L/2$ to $\pi$ at $x=L/2$. The profile of phases and magnetic field are shown in Fig. \ref{f4}.

\begin{figure}[t]
\psfig{figure=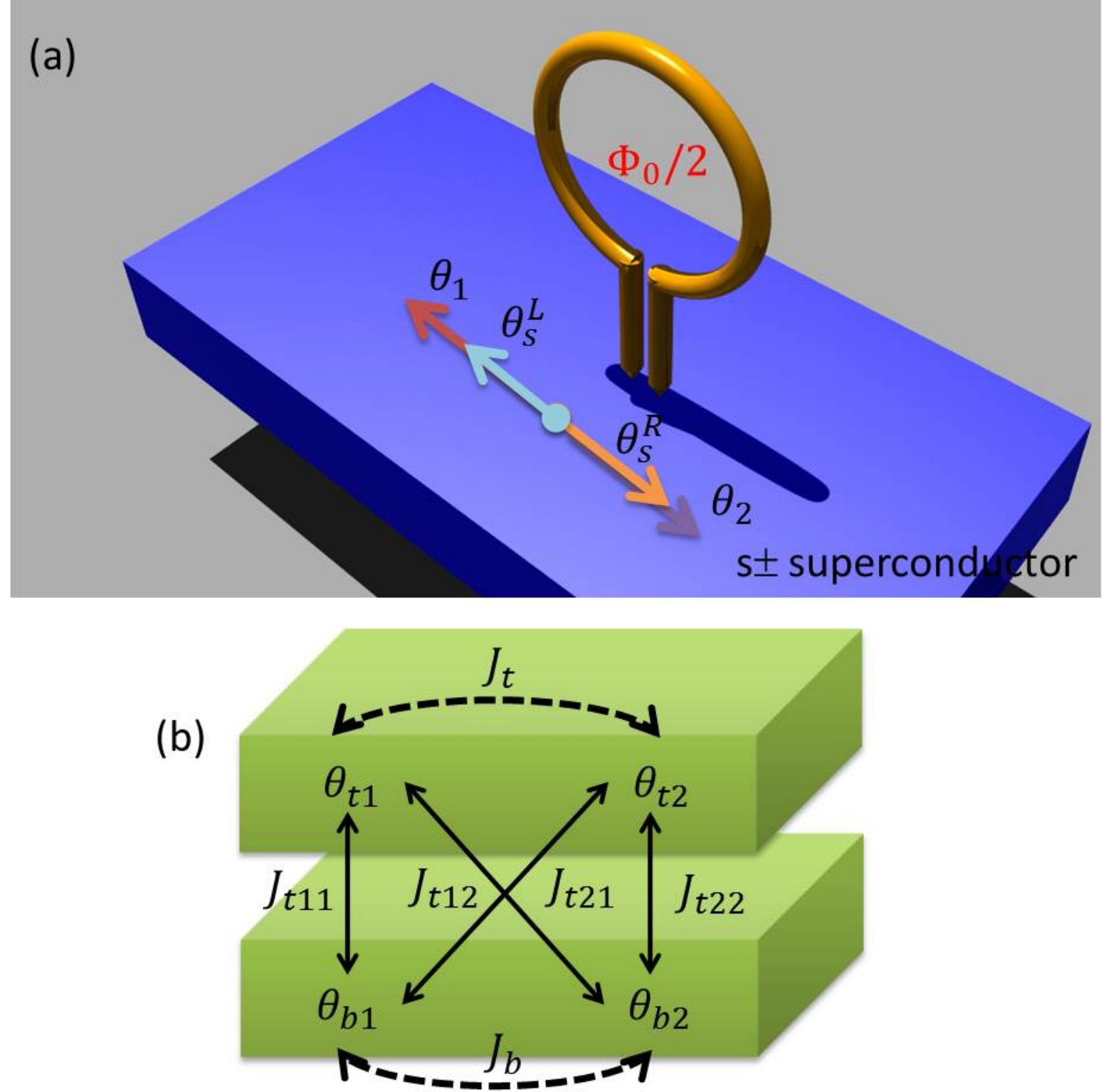,width=\columnwidth} \caption{\label{f10} (color online). (a) Experimental proposal to detect the pairing symmetry in multiband superconductors. (b) Grain boundary Josephson junctions in multiband superconductors.}
\end{figure} 

The modulation of the Josephson coupling Eq. (\ref{eq32}) can be simplified into a setup shown in Fig. \ref{f10}(a), where a loop made of conventional s-wave superconductor is in contact with a two-band superconductor and forms two point junctions. This geometry is the same as the one studied experimentally in Ref. \cite{Chen09}. When $|J_{12}|\gg J_{si}$, $\theta_1=\pi+\theta_2$. If at the left contact $J_{s1}>J_{s2}$, then $\theta_s^L=\theta_1$; However, if one can achieve at the right end $J_{s2}<J_{s1}$, then $\theta_s^R=\theta_2=\theta_1+\pi$ when the self-inductance is small for a large loop. Then there is a $\pi$ phase shift between the two ends of the s-wave superconductor. This is a new way to realize the $\pi$ junction\cite{Bulaevskii77} by inserting a two-band superconductor with $s\pm$ pairing symmetry between two conventional s-wave superconducting electrodes. For a superconducting loop containing a $\pi$ junction, TRSB occurs with a spontaneous magnetic flux of $\Phi_0/2$. If this spontaneous magnetic flux is measured, then it can unequivocally prove the sign-reversal pairing symmetry for the two-band superconductors.

The above analysis can be generalized into Josephson junctions in grain boundaries of $s\pm$ multiband superconductors. As shown in Fig. \ref{f10}(b), there are four tunnelling channels in the grain boundary junction, with two diagonal and two off-diagonal tunnellings. The total energy can be written as
\begin{align}\label{eq39a}
\nonumber E=-J_{t11}\cos(\theta_{t1}-\theta_{b1})-J_{t12}\cos(\theta_{t1}-\theta_{b2})\\
\nonumber-J_{t21}\cos(\theta_{t2}-\theta_{b1})-J_{t22}\cos(\theta_{t2}-\theta_{b2})\\
-J_{t}\cos(\theta_{t1}-\theta_{t2})-J_{b}\cos(\theta_{t1}-\theta_{t2}),
\end{align}
where $J_{t\alpha\beta}>0$ with $\alpha,\ \beta=1\ ,\ 2$ are the inter-junction couplings, and $J_t<0$, $J_b<0$ are inter-band couplings. For $|J_t|$ and $|J_b|\gg J_{t\alpha\beta}>0$, $\theta_{t1}=\theta_{t2}+\pi$ and $\theta_{b1}=\theta_{b2}+\pi$, then the Josephson energy is simplified into
\begin{equation}\label{eq39b}
E=(-J_{t11}-J_{t22}+J_{t12}+J_{t21})\cos(\theta_{t2}-\theta_{b2}).
\end{equation}
If at some part of the junction $(-J_{t11}-J_{t22}+J_{t12}+J_{t21})>0$, while at other part $(-J_{t11}-J_{t22}+J_{t12}+J_{t21})<0$, then we have exactly the same situation as studied in Eqs. (\ref{eq32}) and (\ref{eq33}). Spontaneous magnetic flux will appear in the grain boundary if this condition holds.

\begin{figure}[t]
\psfig{figure=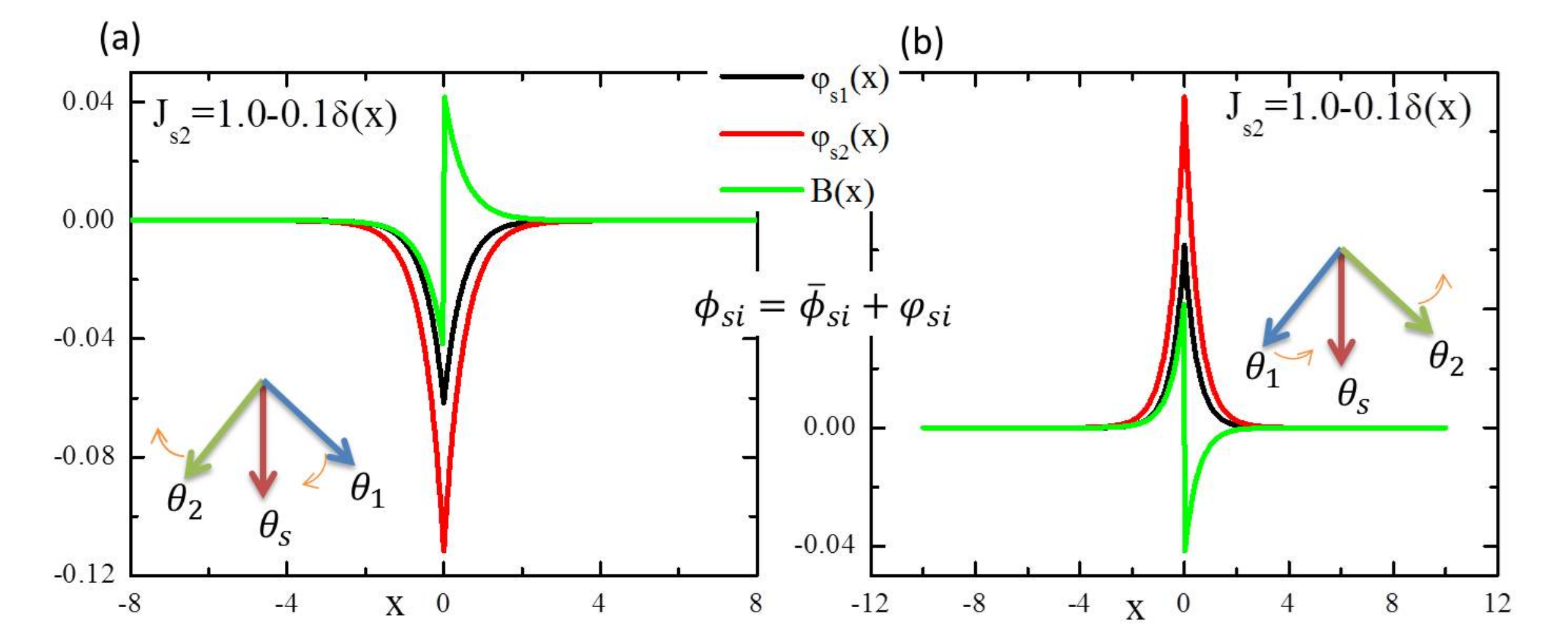,width=\columnwidth} \caption{\label{f5} (color online). Profile of the phases and magnetic field with a defect in the junction with $s\pm$ pairing symmetry obtained from Eqs. (\ref{eq44}-\ref{eq46}). The defect is modelled as inhomogeneous Josephson coupling in the junction. (a) and (b) correspond to two TRSB pair states. Inset is the phase configuration in the ground state.}
\end{figure} 
\begin{figure}[b]
\psfig{figure=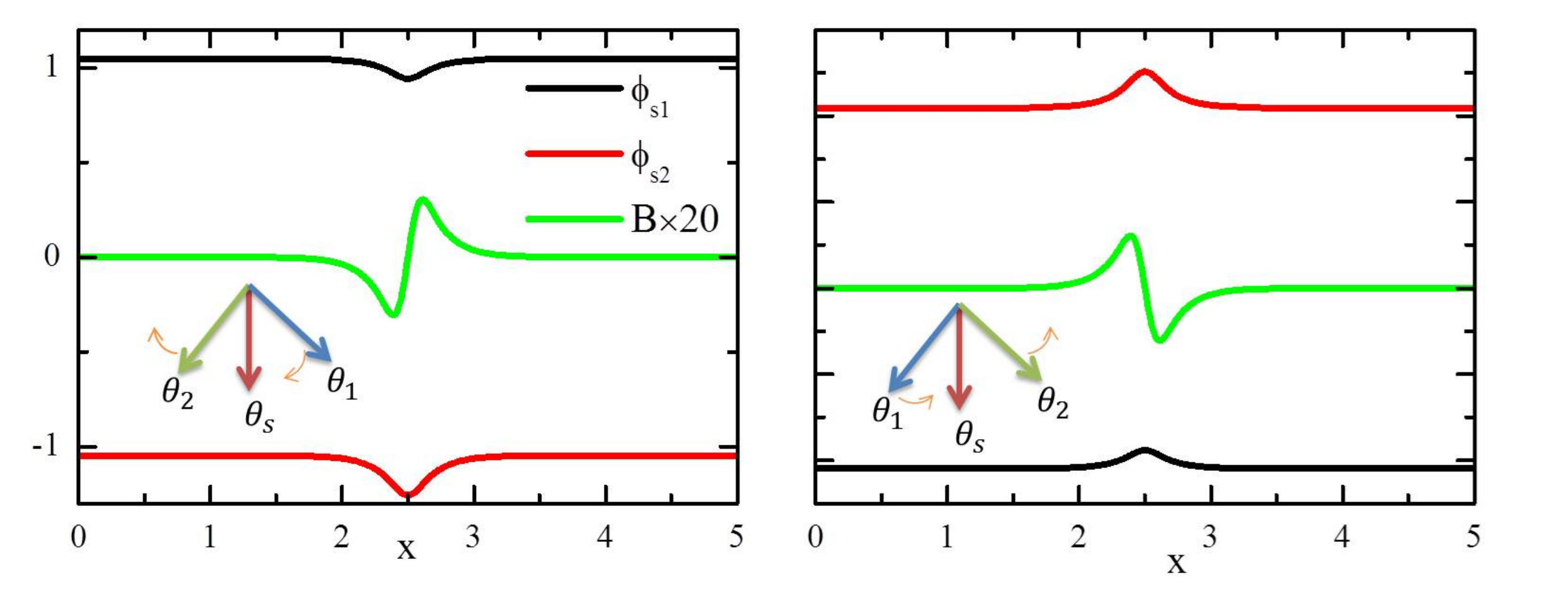,width=\columnwidth} \caption{\label{f6} (color online). Profile of the phases and magnetic field with a Gaussian-type defect $J_{s2}(x)=1-0.4\exp(-(x-L/2)^2)$ in the Josephson coupling in a junction with $s\pm$ pairing symmetry obtained by numerical calculations. (a) and (b) correspond to two TRSB pair states. Inset is the phase configuration in the ground state. For clarity, the magnetic field is amplified by 20 times.}
\end{figure} 

We then investigate the response of a junction in the TRSB state when spatial inhomogeneities of the Josephson couplings are present. In the TRSB state, $\phi_{si}$ depends on the Josephson couplings. The inhomogeneities induce variation of $\phi_{si}$, and thus create magnetic field. We assume a weak defect in the $J_{s2}$ channel $J_{\text{s2}}=\bar{J}_{\text{s2}}+\Delta _{\text{s2}}(x)$, and the response is $\phi _{\text{si}}=\bar{\phi }_{\text{si}}+\varphi _{\text{si}} $ with $\Delta _{\text{s2}}\ll1$ and $\varphi_{si}\ll 1$. For simplicity, we assume two channels are identical $\zeta_1=\zeta_2=\zeta_s=\zeta$, $\bar{J}_{\text{s2}}=J_{s1}$ and $\bar{\phi }_{\text{s2}}=-\bar{\phi }_{\text{s1}}=\phi _0$. Substituting these expressions into Eqs. (\ref{eq17}) and (\ref{eq21}) and then expanding to the linear order, we have
\begin{equation}\label{eq40}
\frac{\partial _x^2\varphi _{\text{s1}}+\partial _x^2\varphi _{\text{s2}}}{3\zeta +2}=\cos {\phi }_{0}\varphi _{\text{s1}}+\Delta _{\text{s2}}(x)\sin {\phi }_{0}+\bar{J}_{\text{s2}}  \varphi _{\text{s2}}\cos {\phi }_{0},
\end{equation}
\begin{align}\label{eq41}
\nonumber-\frac{3\zeta +1}{3\zeta +2}\frac{\partial _x^2\varphi _{\text{s1}}}{\zeta } +\frac{\partial _x^2\varphi _{\text{s2}} }{\zeta(3\zeta +2)}+2\cos {\phi }_{0}\varphi _{\text{s1}}+\Delta _{\text{s2}}(x)\sin {\phi }_{0}\\
+\bar{J}_{\text{s2}} \cos {\phi }_{0} \varphi _{\text{s2}}+J_{12}\cos(2\phi_0) \left(\varphi _{\text{s1}}-\varphi _{\text{s2}}\right)=0.
\end{align}
Equations (\ref{eq40}) and (\ref{eq41}) can be solved in the Fourier space and the solution is given by
\begin{equation}\label{eq42}
\varphi _{\text{s1}}(k)=-\frac{\sin\phi_0 \left[2k^2 (1+\zeta )+p^2(2+3 \zeta )-q^2\zeta\right]}{(k^2+p^2 )(k^2+q^2)}\Delta  (k),
\end{equation}
\begin{equation}\label{eq43}
\varphi _{\text{s2}}(k)=-\frac{\sin\phi_0 \left[2k^2 (1+2 \zeta )+p^2(2+3 \zeta )+q^2\zeta\right]}{2(k^2+p^2 )(k^2+q^2)}\Delta  (k),
\end{equation}
with $p^2=\left[\cos\phi_0+2 \cos \left(2\phi _0\right) J_{12}\right] \zeta$ and $q^2=\cos\phi_0 (2+3 \zeta )$. We assume a point defect $\Delta(x)=\Delta_0 \delta(x)$. Then we obtain the induced modulation of phase in real space
\begin{equation}\label{eq44}
\varphi _{\text{s1}}(x)=-\frac{\Delta _0\sin\phi_0}{4}\left[\frac{ e^{-q|x|} q}{\cos\phi_0}-\frac{ e^{-|x| p} p}{\cos\phi_0+2\cos \left(2\phi _0\right) J_{12}}\right],
\end{equation}
\begin{equation}\label{eq45}
\varphi _{\text{s2}}(x)=-\frac{\Delta _0\sin\phi_0}{4}\left[\frac{ e^{-q|x|} q}{\cos\phi_0}+\frac{ e^{-|x| p} p}{\cos\phi_0+2\cos \left(2\phi _0\right) J_{12}}\right],
\end{equation}
and the associated magnetic flux is given by
\begin{equation}\label{eq46}
B=\frac{2\Delta _0\sin\phi_0\text{sign}[x]}{3\zeta +2}\frac{ e^{-|x|q} q^2}{4 \cos\phi_0}.
\end{equation}
Please note that the magnetic flux is singular at $x=0$ because of the $\delta$ function used for the defects. Since Eqs. (\ref{eq44}-\ref{eq46}) all contain $\sin\phi_0$, the response to the same defect for distinct TRSB state differs by a sign, as shown in Fig. \ref{f5} (a) and (b). This can be understood by looking at the phases of the three condensates. We have the freedom to fix the phase of the s-wave superconductor $\theta_s=0$. As shown in the inset of Fig. \ref{f5}(a), when $J_{s2}$ is suppressed due to the local defect, the attraction between $\theta_s$ and $\theta_2$ decreases, thus $\theta_2$ decreases (rotates leftwards). As a result $\theta_1$ also decreases because the repulsion between $\theta_1$ and $\theta_2$ is reduced. The width of the region of phase variation is optimized by energy cost due to the phase gradient. While for the other TRSB ground state shown in Fig. \ref{f5}(b), when $J_{s2}$ decreases, $\theta_2$ increases (rotates rightwards) as a result of the reduced attraction. Meanwhile $\theta_1$ also increases.

For a strong defect, we calculate the phases and magnetic field numerically. We model the point defect by a Gaussian distribution, and the results are presented in Fig. \ref{f6}. Magnetic flux is induced near the defect but the integrated magnetic flux in the junction vanishes. Meanwhile the response for two distinct TRSB states is different, consistent with the analytical results in Eqs. (\ref{eq44}-\ref{eq46}).

\section{Conclusions}
In this work, we have studied systematically the Josephson effect between a two-band superconductor and a conventional $s$-wave superconductor. We consider both the $s++$ and $s\pm$ pairing symmetries for the two-band superconductor. Due to the multiband nature of the superconducting electrode, there are two tunnelling channels in the junction, which gives rise to complicated interference in physical quantities, depending on the underlying pairing symmetry. Moreover for junctions with $s\pm$ pairing symmetry, there exists frustrated interaction among different condensates and under appropriate conditions, time-reversal symmetry is broken. Depending on the competition of inter-junction Josephson couplings and inter-band Josephson coupling, the interference between the two tunnelling channels can change continuously from adding constructively where they have the same phase to cancelling destructively where they have a $\pi$ phase shift. The interference manifests itself in the critical current, Shapiro steps, Fiske current steps and Eck current steps.

In the case of thin superconducting electrodes, charge neutrality can be broken in the electrodes. Because of the charge neutrality breaking, out-of-phase oscillations of the gauge-invariant phase difference between two channels are possible, which gives a new plasma mode, in addition to the in-phase plasma oscillations. The energy gap of the out-of-phase mode vanishes at the time-reversal symmetry breaking transition, which is a first example of massless plasma mode in superconductors. Because of the existence of two plasma modes, there are additional Fiske and Eck resonances in the Josephson flux flow region.     

A long junction supports topological excitation of solitons. When the gauge invariant phase difference associated with one tunnelling channel changes by multiple $2\pi$ that is different from the change of phase difference in the other channel, fractional quantized magnetic solitons are stabilized in the junction. In contrast to the fractional vortices in bulk multiband superconductors, the fractional solitons have finite energy thus are thermodynamically stable. For junctions with time-reversal symmetry breaking, a new type of soliton excitations can be created between two distinct time-reversal pair states.    

Finally we showed that in junctions with time-reversal symmetry breaking, disorders induce magnetic flux in the junctions, which points a unique way to detect the pairing symmetry of multiband superconductors.

\section{Acknowledgement}
The author is grateful to L. N. Bulaevskii for helpful discussions. This work was supported by the US Department of Energy, Office of Basic Energy Sciences, Division of Materials Sciences and Engineering.

%

\end{document}